\pdfoutput=1
\documentclass[pra,twocolumn,showpacs,floatfix]{revtex4-1}
\usepackage{times,amsmath,amssymb,amstext,latexsym,float,graphicx,color,ulem}
\usepackage{hyperref}

\hypersetup{colorlinks=true, citecolor=blue, urlcolor=blue, linkcolor=blue}
\begin{document}

\title{From spontaneous to explicit symmetry breaking in a finite-sized system: \\ Bosonic bound states of an impurity}

\author{L.~Chergui$^1$}
\email{lila.chergui@matfys.lth.se}
\author{F. Brauneis$^2$}
\author{T. Arnone Cardinale$^1$}
\author{M. Schubert$^1$}
\author{A. G. Volosniev$^3$}
\author{S.~M.~Reimann$^1$}

\affiliation{$^1$Division of Mathematical Physics and NanoLund, Lund University, SE-221 00 Lund, Sweden}
\affiliation{$^2$Technische Universität Darmstadt, Department of Physics, 64289 Darmstadt, Germany}
\affiliation{$^3$Department of Physics and Astronomy, Aarhus University, Ny Munkegade 120, DK-8000 Aarhus C, Denmark}

\begin{abstract}
{
The presence of a single attractive impurity in an ultracold repulsive bosonic system can drive a transition from a homogeneous to a localized state, as we here show for a one-dimensional ring system. In the few-body limit the localization of the bosons around the impurity, as seen in the pair correlations, is accompanied by low-lying modes that resemble finite-size precursors of Higgs-Anderson and Nambu-Goldstone like modes. Tuning the impurity-boson mass ratio allows for the exploration of the transition from a spontaneous to an explicit breaking of the continuous rotational symmetry of the Hamiltonian. We compare the minimum of the Higgs-Anderson like mode as a marker of the onset of localization in the few-body limit to mean-field predictions of binding. We find improved agreement between the few-body exact diagonalization results and mean-field predictions of binding with increasing boson-boson repulsion.
}
\end{abstract}

\maketitle

\section{Introduction}

Motivated by theoretical~\cite{PhysRevLett.115.155302,PhysRevLett.117.100401} and experimental~\cite{doi:10.1126/science.aao5686,PhysRevLett.120.235301,PhysRevLett.120.135301} work on quantum liquid droplets in the thermodynamic limit, we previously investigated the formation of self-binding quantum droplets in 1D in the few-body regime~\cite{Chergui_2023}. 
One-dimensional systems are of particular interest~\cite{PhysRevA.101.051601,PhysRevResearch.3.043128,10.21468/SciPostPhys.16.3.074} due to the enhancement of quantum correlation effects in low dimensions~\cite{Sowinski_2019,MISTAKIDIS20231}.
For a two-component bosonic system with equal particle numbers, equal intra-species repulsion in both components and inter-species attraction
we found in the excitation energy spectra, signatures of spontaneous symmetry breaking suggestive of a few-body crossover from a homogeneous to a localized state. The corresponding modes resembled precursors of
Higgs-Anderson (HA)-like amplitude modes~\cite{Higgs1964,Anderson1958} and Nambu-Goldstone (NG)-like rotational modes~\cite{Goldstone1961,PhysRev.127.965,PhysRev.117.648,PhysRev.122.345}.
Similar modes indicating symmetry breaking and an associated phase transition  have been observed in other systems in the limit of large particle numbers, including transitions to solitonic~\cite{Ueda_HA_GS_PhysRevLett.94.090404} and supersolid~\cite{hertkorn2019fate} states in Bose-Einstein condensates (BEC).
In the few-to many-body regime ~\cite{Bjerlin2016,Bayha2020, PhysRevA.110.L061302} similar mode precursors were also discussed in the context of paired fermions in two dimensions.

Here, we study a maximally unbalanced two-component bosonic system -- the case of a single attractive impurity in the presence of repulsive bosons. 
We find similar signatures of symmetry breaking due to a few-body phase crossover, manifest in the pair correlations and in the low lying energy spectra. These features were also found in the symmetric two-component bosonic system~\cite{Chergui_2023}, indicating that they are more general nature.

In the following we analyze the exact few-body spectra of the impurity system and show that 
the position of the minimum of the first excitation in the ${L = 0}$ subspace, ${g_{ab}^*}$, can be identified as a marker of the onset of localization in the cross-over region from  a homogeneous to a localized state.
This work is organized as follows, after this introduction Section~\ref{Sec: Model} provides a brief description of the model used. In Sec.~\ref{localization and few-body crossover region} we examine the connection between the low-lying excitations and localization in the pair correlations via the Hellmann-Feynman theorem.
We then compare the condition of the minima of the HA-like modes 
to  the critical condition for the binding of repulsively interacting bosons to a single attractive impurity derived for a one-dimensional system in  the mean-field (MF) limit in Ref.~\cite{Brauneis_2022}.
While the MF condition for localization is independent of the boson-impurity mass ratio, this ratio remains a degree of freedom of the Hamiltonian for the few-body system considered in this work. 
In the comparison with MF predictions we present exact diagonalization results for the position of the minimum of the HA-like mode of the few-body system for various impurity masses.
In Sec.~\ref{Sec: Impurity-Boson Mass Ratio} we investigate the effect of the impurity mass on the ground state energy as a function of total angular momentum, and study how it drives a transition from a spontaneous to an explicit breaking of the rotational symmetry of the system. 
The Hamiltonian describing a system of bosons with finite mass interacting via contact interactions on a one-dimensional ring is invariant under the transformation of an infinitesimal rotation. Therefore, it has a continuous rotational symmetry (see e.g.~\cite{Sakurai_Napolitano_2020}). If the ground state of such a system does not respect this symmetry, for example through localization in the ground state density, it is said to be spontaneously broken. If the symmetry of the Hamiltonian is broken by the inclusion of an additional term, such as a potential with a fixed position on the ring, corresponding to an infinitely massive impurity, then the symmetry is explicitly broken.
We calculate the excitation energies for various impurity masses and 
demonstrate how for increasing impurity mass the HA-like mode of the system with spontaneous symmetry breaking converges towards the result for an infinitely massive impurity, {i.e.}, a fixed delta potential.
We interpret this as a transition from spontaneous to explicit symmetry breaking.
Finally, a summary is given in Sec.~\ref{Sec: Summary} along with an outlook to future prospective investigations.
In Appendix~\ref{appendix: Delta Potential Trap} we provide details of the exact diagonalization calculations for the infinitely massive impurity limit.
Appendix~\ref{Appendix: Convergence} provides further details of the exact diagonalization calculations for the finite mass impurity systems, including convergence data.
In Appendix~\ref{Appendix - PC Hellmann-Feynman} we show that at the minimum of the first excitation mode in the zero angular momentum subspace, the Hellmann-Feynman theorem~\cite{Hell-Feynman-PhysRev.56.340} implies the equality of the pair correlations of the bosons in the ground state and first excited state at the fixed position of the impurity. Appendix~\ref{Appendix: BdG} details the calculations of the BdG excitation energies.

\section{Model}\label{Sec: Model}
We consider a system of $N_b$ identical, repulsive bosons of mass $M_b$ in the presence of a single attractive impurity of mass $M_a$ confined to a one-dimensional ring of radius $R$. In ultracold atomic systems, the interactions are well described by the contact interaction (see {e.g.} Ref.~\cite{RevModPhys.80.885}) and the Hamiltonian is given by
\begin{equation}
    \begin{split}
       \hat{H} = &-\frac{\hbar^2}{2 M_a R^2}\frac{\partial^2}{\partial\theta_{a}^2} - \frac{\hbar^2}{2 M_b R^2}\sum_{i=1}^{N_b} \frac{\partial^2}{\partial \theta_{b,i}^2} \\
       &+ \frac{g_{bb}}{R}\sum_{i>j}\delta(\theta_{b,i} - \theta_{b,j}) + \frac{g_{ab}}{R}\sum_{i=1}^{N_b}\delta(\theta_{b,i} - \theta_a ). 
    \end{split} 
     \label{eqn: hamiltonian}
\end{equation}
Here, $g_{bb} > 0$ and $g_{ab} < 0$ are the strengths of the boson-boson repulsion and boson-impurity attraction respectively. We present results in units defined by ${\hbar = R = M_b =1}$. The impurity-boson mass ratio ${M_a/M_b = M_a}$ is varied as a parameter of the system.

For few-body systems such as we consider here, the Hamiltonian Eq.~\eqref{eqn: hamiltonian} can be diagonalized via exact diagonalization/configuration interaction (CI) methods. A key advantage of the CI method is that in addition to the ground state, it gives access to the energies and wave functions of excitations of the system, which are central to the analysis of this work. We employ a similar procedure as outlined in Ref.~\cite{Chergui_2023} wherein the one-body basis consists of the angular momentum eigenstates
\begin{equation}
    \phi_m(\theta) = \frac{1}{\sqrt{2\pi}}e^{im\theta}
    \label{eqn: one-body basis}
\end{equation}
with integer one-body angular momentum quantum numbers $|m| \leq m_\textrm{Max}$. We utilise the block diagonal structure of the Hamiltonian due to the conservation of total angular momentum, $L$. An importance-truncation scheme~\cite{PhysRevC.79.064324,Tubman2020} is employed to search the resulting Hilbert space for the subset of many-body states that contribute to the target eigenstate of Eq.~\eqref{eqn: hamiltonian}.
The importance-truncated configuration interaction (ITCI) method allows for access to large one-body angular momentum cut-offs, e.g. ${m_\textrm{Max} = 60}$, required to capture the comparatively large interaction energies and highly correlated nature of the states of the localized regime. For details of such methods, see e.g. Refs.~\cite{PhysRevC.79.064324,Tubman2020}.
In addition to the ITCI method, we employ running coupling constants to further aid in convergence as described in Ref.~\cite{brauneis2024}.

In the limit ${M_a\rightarrow\infty}$ the impurity acts as a fixed delta potential, which can be treated as an external potential on the ring. 
In this case, the Hamiltonian is obtained from Eq.~\eqref{eqn: hamiltonian} by neglecting the kinetic energy of the impurity and fixing its position ${(\theta_a = 0)}$. For details of the exact diagonalization calculations in this limit including all relevant convergence parameters, see Appendix~\ref{appendix: Delta Potential Trap}. 
For convergence data for the finite mass impurity systems see Appendix~\ref{Appendix: Convergence}.

\section{localization and the few-body crossover region}\label{localization and few-body crossover region}

\subsection{ Pair Correlations and the Hellmann-Feynman Theorem}\label{Sec: pair correlations and hellmann-feynman}

\begin{figure}[hpt!]
    \centering
    \includegraphics[width=\columnwidth]{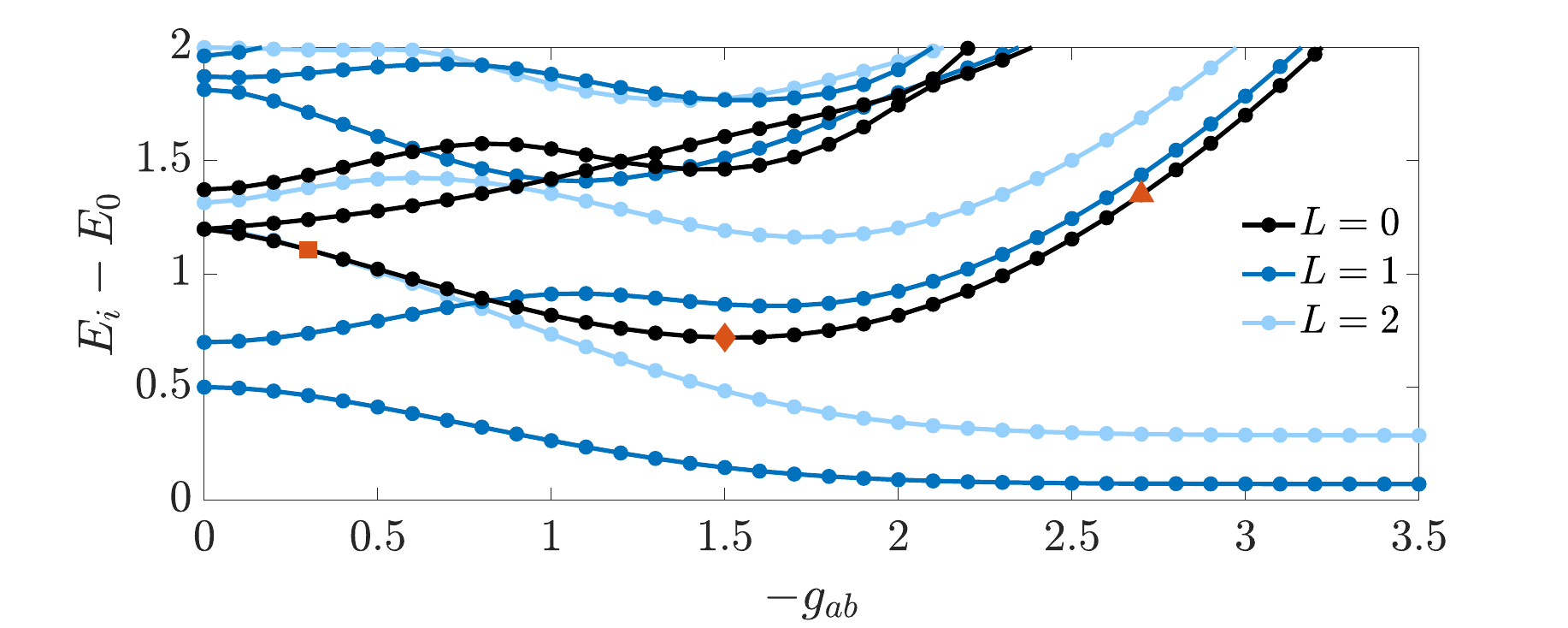}
    \includegraphics[width=\columnwidth]{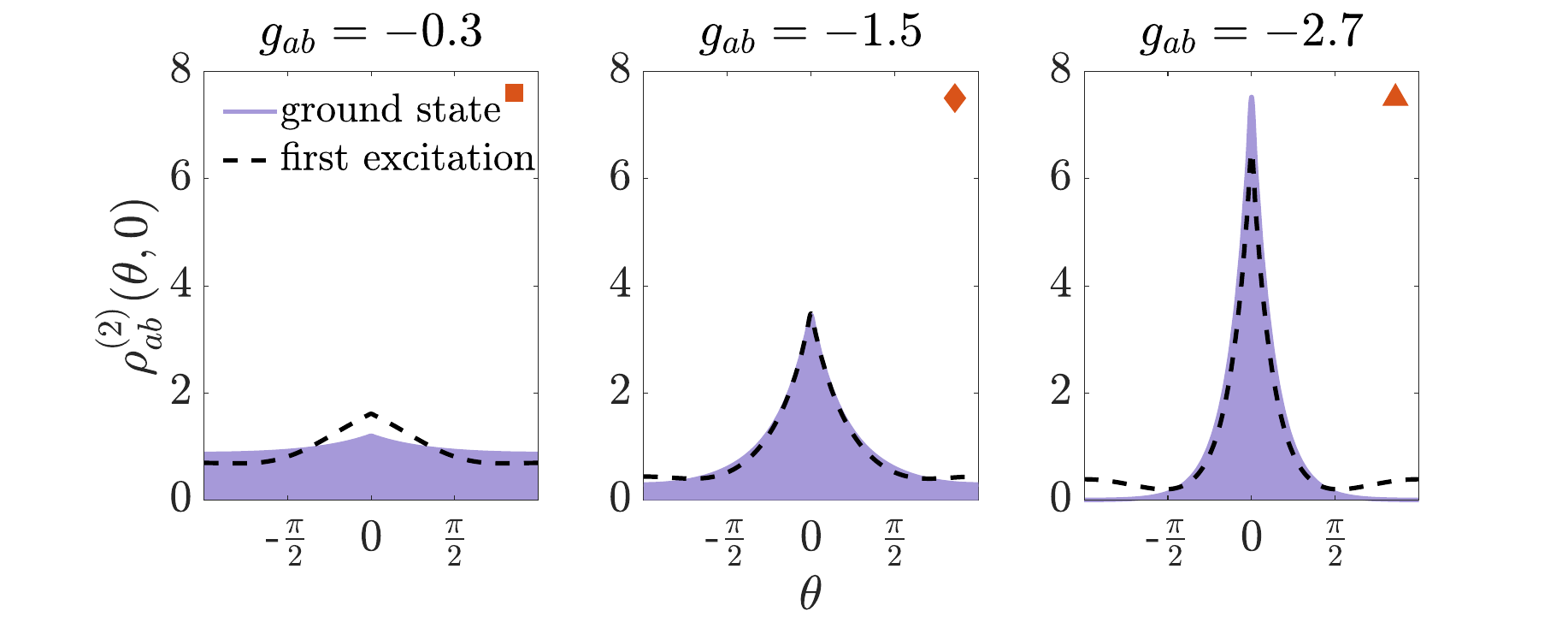}
    \caption{(Top) Excitation energy spectrum as a function of boson-impurity attraction for a system of ${N_b = 6}$, ${g_{bb} = 0.3}$, and ${M_a = 1.0}$. The excitation energies are plotted in black, blue and light blue for the {$L = 0, 1,$ and $2$} subspaces respectively. Circular markers indicate calculated points and connecting lines are included to guide the eye. (Bottom) Pair correlations for the ground state (solid purple) and first excited state (dashed line) with {$L = 0$} for a system of ${N_b = 6}$, ${g_{bb} = 0.3}$, and ${M_a = 1.0}$. The square, diamond and triangle markers indicate the associated excitations.} 
    \label{fig: pair correlations N = 6 g = 0.3}
    \centering
\end{figure}

Let us now consider a system of ${N_b = 6}$ repulsive ${(g_{bb} = 0.3)}$ bosons and a single impurity with mass ${M_a = 1}$.
In Fig.~\ref{fig: pair correlations N = 6 g = 0.3} (top) we show the excitation energy spectrum of the system as a function of boson-impurity attraction.
As in the symmetric case~\cite{Chergui_2023},
in this unbalanced system we see in the ${L = 0}$ subspace (black) the formation of a distinct minimum in its first excitation.  
Furthermore, in the ${L \neq 0}$ subspaces (blue, light blue) the lowest excitations monotonically decrease and asymptotically approach ${L^2}/{2(N_bM_b + M_a)}$, the energy of a rigid body of mass $N_bM_b + M_a$ rotating with angular momentum $L$ around the ring. In the limit of large $N_b$ these rotational modes will become degenerate with the ground state, suggesting that these modes are few-body precursors to NG-like phase modes. These modes indicate a spontaneous transition to a localized state as they represent the rigid body rotation of the localized ground state around the ring. 

The transition to a localized state can also be seen in the pair correlations
\begin{equation}\label{eqn: pair correlations}
    \begin{split}
        \rho^{(2,i)}_{\sigma\sigma'}(\theta,\theta') = \sum_{m,n,k,l}\phi_m^*&(\theta)\phi_n^*(\theta')\phi_{k}(\theta')\phi_{l}(\theta) \\
        &\times\langle \psi_i| \hat{a}^\dagger_{\sigma,m}\hat{a}^\dagger_{\sigma',n}\hat{a}_{\sigma',k}\hat{a}_{\sigma,l}|\psi_i\rangle
    \end{split}
\end{equation}
for component ${\sigma \in \{a,b\}}$ with respect to the position $\theta '$ of a single fixed particle of component ${\sigma ' \in \{a,b\} }$ for the system in the state ${|\psi_i\rangle}$. Here ${\hat{a}_{\sigma,m}^\dagger}$ (${\hat{a}_{\sigma,m}}$) is the creation (annihilation) operator for species $\sigma$ in the single-particle angular momentum eigenstate Eq.~\eqref{eqn: one-body basis}.
Figure~\ref{fig: pair correlations N = 6 g = 0.3} (bottom) shows the pair correlations of the bosons ${(\sigma = b)}$ with respect to the fixed position of the impurity ${(\sigma' = a, \theta' = 0)}$ 
for the ground state {(solid purple)} and first excited state {(dashed black line)} of the system in the ${L = 0}$ subspace 
with ${g_{ab} = -0.3, -1.5}$, and ${g_{ab} = -2.7}$ (left to right). With stronger inter-component attraction, $g_{ab}$, we see a transition from a homogeneous to a localized ground state, and a similar profile in the first excitation.

By comparing the pair correlations of the ground state and first excited state we can gain some insight into the behavior of the HA-like mode. For weak boson-impurity attraction the ground state pair correlations of the bosons are approximately homogeneous due to the boson-boson repulsion, and the first excited state pair correlations have a sinusoidal contribution.
%}
This peaking of the boson pair correlations in the first excited state allows for greater overlap between the bosons and the impurity, increasing the relative (negative) contribution of the boson-impurity interaction to the energy of this state. With increasing $-g_{ab}$ this effect becomes more pronounced and the first excitation energy decreases until the mode reaches its minimum, $-g_{ab}^*$. From Fig.~\ref{fig: pair correlations N = 6 g = 0.3} (lower panel) this appears to coincide with the point of largest overlap between the ground state and first excited state pair correlations. As the boson-impurity attraction is further increased the first excited state cannot be more localized than the ground state while still having higher energy. We see this in the pair correlations of the first excited state for $g_{ab} = -2.7$ which deviate from the ground state pair correlations primarily by an increased probability of finding a boson further away from the position of the impurity, as compared to the ground state. This is consistent with the interpretation of the HA-like mode as a breathing mode excitation of the ground state~\footnote{In order for the superposition of the ground state and first excitation to “breath”, i.e. for the mean angular width of the pair correlations of the superposition state to oscillate periodically in time, the excitation should have less localized pair correlations than the ground state.}, as was found in the case of a symmetric two-component bosonic system~\cite{Chergui_2023}.

We can quantify the relationship between the signatures of localization in the low-lying energy spectra and in the pair correlations via the Hellmann-Feynman theorem~\cite{Hell-Feynman-PhysRev.56.340},
\begin{equation}
    \frac{dE^{(\lambda)}}{d\lambda} = \langle \psi^{(\lambda)} |\frac{d\hat{H}^{(\lambda)}}{d\lambda}|\psi^{(\lambda)}\rangle.
    \label{eqn: Hell-Feynman}
\end{equation}
Here $E^{(\lambda)}$ and $\psi^{(\lambda)}$ are an eigenvalue and the corresponding eigenstate of the Hamiltonian $\hat{H}^{(\lambda)}$ with parameter $\lambda$. The position of the minimum of the first ${L=0}$ excitation is simply determined by the equality of the first derivative with respect to $g_{ab}$ of the energies of the ground state and first excited state. Then it follows from Eq.~\eqref{eqn: Hell-Feynman} that the relation
\begin{equation}
    \langle{\psi_0|\frac{\partial \hat{H}}{\partial g_{ab}}|\psi_0}\rangle = \langle{\psi_1|\frac{\partial \hat{H}}{\partial g_{ab}}|\psi_1\rangle}
    \label{eqn: expectation of Hamiltonian derivative}
\end{equation}
holds at the position of the minimum, which we denote $g_{ab}^*$. We show in Appendix~\ref{Appendix - PC Hellmann-Feynman} that this condition is equivalent to the equality of the pair correlations of the bosons in the ground state and first excited state at the fixed position of the impurity, 
\begin{equation}
    \rho^{(2,0)}_{ba}(0,0) = \rho^{(2,1)}_{ba}(0,0).
    \label{eqn: equality of pair correlations}
\end{equation}
The pair correlations of the bosons with respect to a fixed impurity are equivalent to the density of the bosons in the co-moving frame of the impurity,
\begin{equation}
    |\psi_0(0)|^2 = |\psi_1(0)|^2.
\end{equation}
Without loss of generality, we have taken the position of the impurity to be $\theta' = 0$.

Returning to Fig.~\ref{fig: pair correlations N = 6 g = 0.3} we see that in addition to the onset of localization in the pair correlations, Eq.~\eqref{eqn: equality of pair correlations} is indeed satisfied at ${g_{ab}^* = -1.5}$.
We note that Eq.~\eqref{eqn: expectation of Hamiltonian derivative} holds for arbitrary particle numbers and is not particular to the case of a single impurity. It is directly applicable to systems with a HA-like mode indicating a spontaneous symmetry breaking driven by a free parameter of the Hamiltonian. Eq.~\eqref{eqn: equality of pair correlations} is applicable to systems where the phase transition or few-body phase crossover may be driven by the strength of a delta-potential interaction. 

\subsection{Comparison to Critical Binding Condition}\label{Sec: Comparison to Mean-Field}

We now compare the position of the HA-like minima, $g_{ab}^*$, for various systems described by the Hamiltonian, Eq.~\eqref{eqn: hamiltonian} to MF predictions of the onset of localization.
In Ref.~\cite{Brauneis_2022}, conditions for the binding of repulsively interacting bosons to a single attractive impurity were studied for a one-dimensional system. The authors found that the condition which differentiates scattering state solutions from solutions in which all bosons are bound to the impurity is given by
\begin{equation} 
    N_{cr}=2\frac{|g_{ab}|}{g_{bb}}+1.
    \label{eqn: MF Ncrit condition}
\end{equation}
For particle numbers ${N_b \leq N_{cr}}$ all bosons are bound to the impurity, {i.e.}, the probability to measure a boson far away from the impurity vanishes. If the particle number is larger, ${N_b > N_{cr}}$, some bosons can be found in scattering states. Note that Eq.~\eqref{eqn: MF Ncrit condition} is independent of the boson and impurity masses. These results were obtained with a MF ansatz in the frame co-moving with the impurity for an infinite one-dimensional line, see also Ref.~\cite{PhysRevA.69.063401,J_C_Gunn_1988}.  

\begin{figure}[hpt]
    \centering
    \includegraphics[width=\columnwidth]{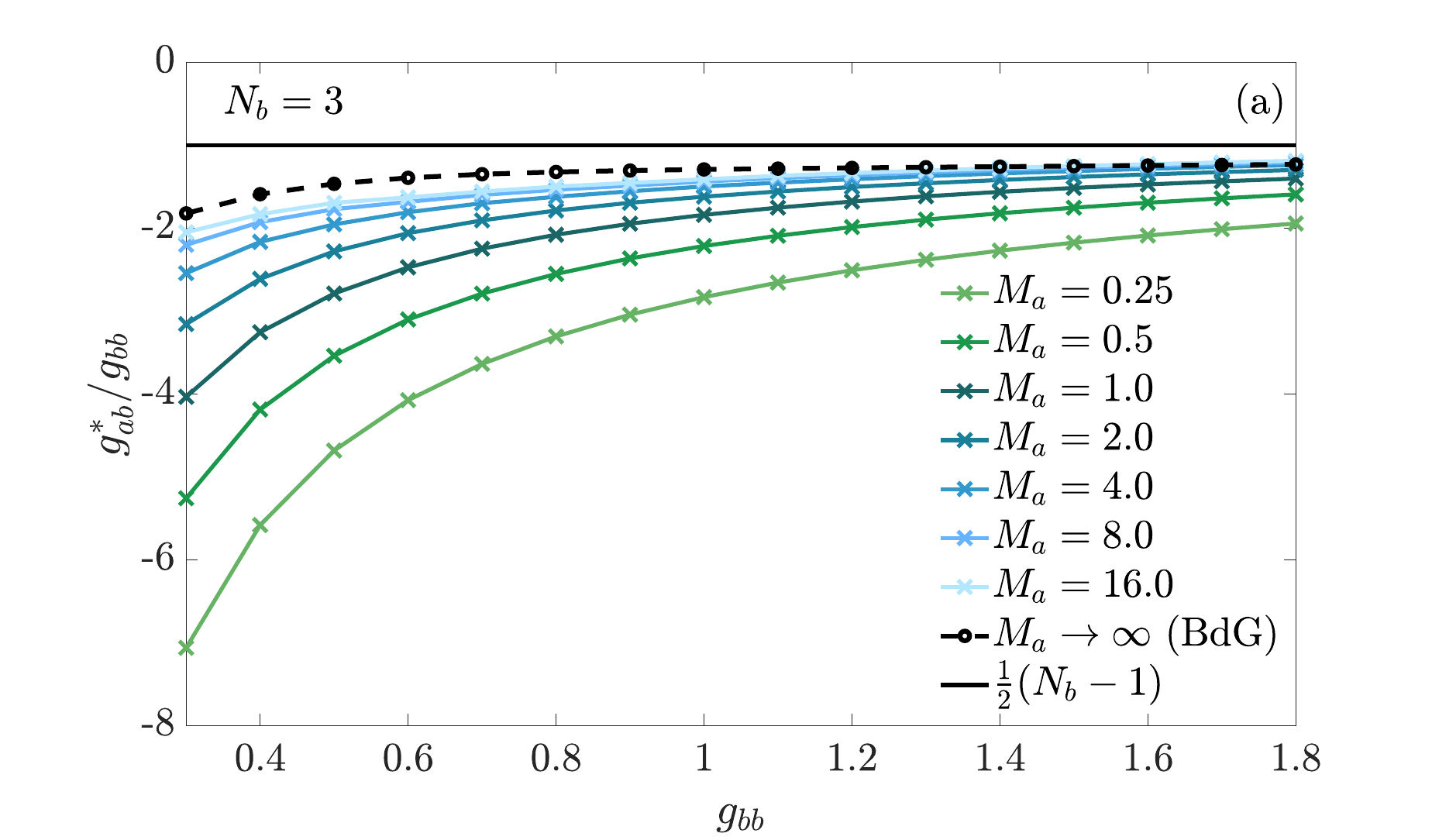}
    \includegraphics[width=\columnwidth]{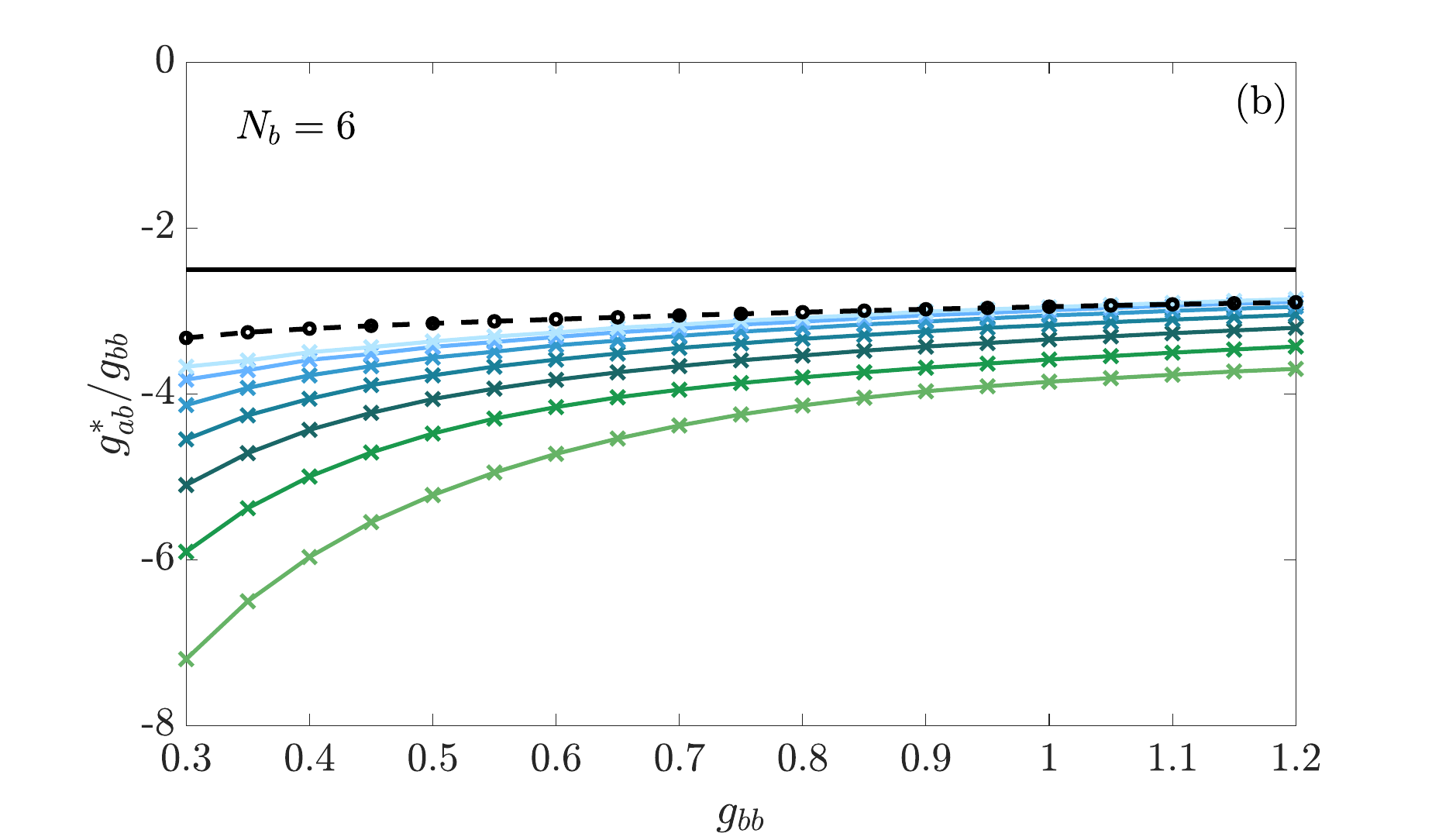}
    \caption{Position of the first ${L = 0}$ excitation minimum at impurity-boson interaction strength $g_{ab}^*$, relative to the boson-boson repulsion, $g_{bb}$, as a function of $g_{bb}$ for a system of ${N_b =  3}$ (top) and ${N_b = 6}$ (bottom) bosons and various impurity masses, $M_a$. The solid black line indicates ${\frac{1}{2}(N_b - 1)}$, the MF prediction of the position of the phase transition from Eq.~\eqref{eqn: MF Ncrit condition}~\cite{Brauneis_2022}. The exact diagonalization results are indicated by the cross markers with connecting lines to guide the eye. For each marker the HA-like mode is computed as a function of ${-g_{ab}}$ in steps of 
    {at most $0.3$}
    %$0.3$ (or $0.1$ for the modes with ${M_a  = 8, 16}$ for ${N_b = 6}$ and ${g_{bb}} = 1.05-1.2$) 
    and the minima is found with spline interpolation.
    The dot-dashed line indicates the position of the HA minima found from the BdG excitation to the numerical ground state solution of the corresponding GPE for a system of $N_b$ bosons with an infinitely massive (fixed delta potential) impurity. } 
    \label{fig: MF Comparison}
    \centering
\end{figure}

By fixing ${N_\textrm{cr} = N_b}$ Eq.~\eqref{eqn: MF Ncrit condition} provides a MF prediction for $g_{ab}^*$ as a function of $g_{bb}$.
In Fig.~\ref{fig: MF Comparison} the positions of ${g_{ab}^*/g_{bb}}$ are plotted as a function of ${g_{bb}}$ for various impurity masses. We take ${N_\textrm{cr} = N_b = 3, 6}$ (top, bottom) and compare the exact diagonalization results for ${g_{ab}^*}/g_{bb}$
to Eq.~\eqref{eqn: MF Ncrit condition}
(black line). The HA-like modes are calculated as a function of $g_{ab}$ in steps of {at most $0.3$}
and the minima are found with spline interpolation. 
The dot-dashed line indicates the position of the minima of the first excitation energy found from the Bogoliubov de Gennes (BdG) excitation to the numerical ground state solution of the corresponding Gross-Pitaevskii equation (GPE) with an infinitely massive (fixed delta potential) impurity. 
For details of the calculations of the BdG excitations see Appendix~\ref{Appendix: BdG}.
We see immediately that for weak $g_{bb}$ there is no good agreement between the MF prediction and the few-body results. For increasing $g_{bb}$ and increasing $M_a$
the few-body results tend towards the MF prediction.
Indeed Eq.~\eqref{eqn: MF Ncrit condition} appears to provide an upper limit on $g_{ab}^*$ for a system of ${N_{b}=N_\textrm{cr}}$ bosons with a given $g_{bb}$.
Simultaneously increasing $g_{bb}$ and $g_{ab}$ by the same factor is equivalent to increasing the radius of the ring, $R$, or decreasing the density for a system with fixed particle numbers, as can bee seen from Eq.~\eqref{eqn: hamiltonian}. 
Therefore, in the large $R$ limit
we observe the greatest similarity between the MF critical binding condition and the condition of the minima of the HA-like mode.
Finally, we note that the mass dependence of the curves in Fig.~\ref{fig: MF Comparison} allows us to estimate the strength of beyond mean-field effects. Indeed, according to mean-field calculations of Ref.~\cite{Brauneis_2022}, the energy of the system with an impurity of finite mass can be related to the energy of the system with a static impurity as follows
\begin{equation}\label{eqn: energy scaling with mu}
    E(M_a,g_{ab},g_{bb}) = E({M_a \rightarrow\infty}, \mu g_{ab}, \mu g_{bb})/\mu
\end{equation}
where $\mu = M_a/(M_a + 1)$ is the reduced mass. We demonstrate departure from this scaling in Appendix~\ref{Appendix: Reduced Mass Scaling}, which indicates that beyond-mean-field effects are important and necessitates the use of CI methods.

\section{Mass Imbalanced Systems }\label{Sec: Impurity-Boson Mass Ratio}

\subsection{Spontaneous to Explicit Symmetry Breaking}

\begin{figure}[t!]
    \centering
    \includegraphics[width=\columnwidth]{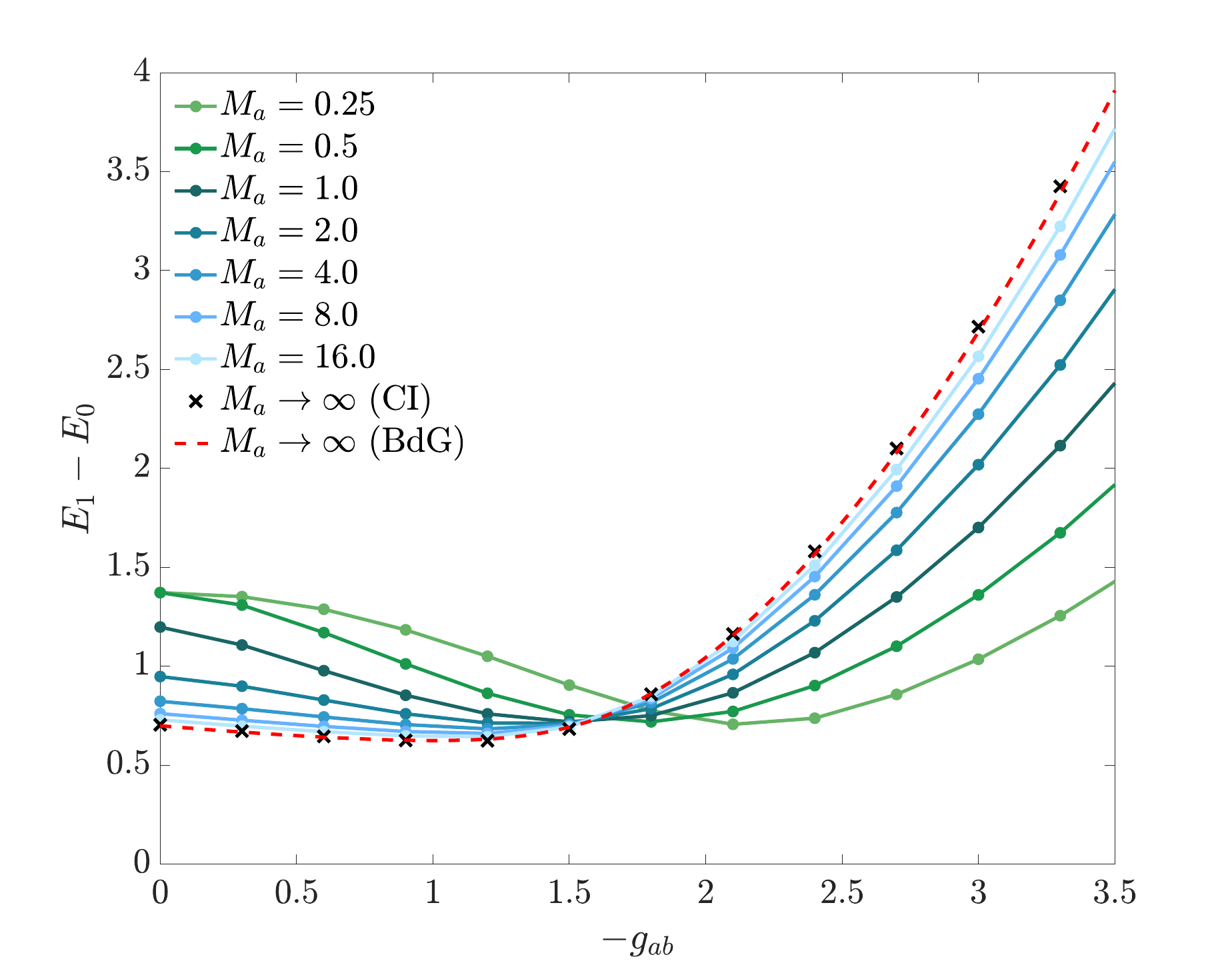}
    \caption{First ${L = 0}$ excitation as a function of boson-impurity interaction strength, $g_{ab}$, for a system of ${N_b = 6}$, ${g_{bb} = 0.3}$, and various impurity masses, ${0.25 \leq M_a \leq 16.0}$. The first excitation for the Hamiltonian with an explicitly broken rotational symmetry due to the fixed delta potential $(M_a \rightarrow \infty)$, calculated in the CI framework described in Appendix~\ref{appendix: Delta Potential Trap}, is plotted in black crosses. The red dashed line shows the first BdG excitation of the numerical ground state solution to the associated GPE.} 
    \label{fig: first excitation for N = 6 g = 0.3 and various M_a }
    \centering
\end{figure}

The Hamiltonian, Eq.~\eqref{eqn: hamiltonian}, has an additional degree of freedom, the impurity-boson mass ratio $M_a/M_b$, or $M_a$ in units defined by $M_b = 1$, which we now consider in greater detail.
Fig.~\ref{fig: first excitation for N = 6 g = 0.3 and various M_a } presents the first excitation in the ${L=0}$ subspace as a function of $-g_{ab}$ for a system of ${N_b=6}$ bosons with ${g_{bb} = 0.3}$ and various impurity masses ${0.25 \leq M_a \leq 16.0}$.
Additionally, the first excitation for the system where the rotational symmetry of the Hamiltonian is explicitly broken by a fixed delta potential ${(M_a \rightarrow \infty)}$ is plotted. Black markers indicate the points calculated in the CI approach and the red dashed line is the first BdG excitation of the numerical ground state solution to the corresponding GPE equation. The agreement between the CI and MF results is noteworthy.
(For details of the CI calculations for an explicitly broken rotational symmetry see Appendix~\ref{appendix: Delta Potential Trap}.)

From Fig.~\ref{fig: first excitation for N = 6 g = 0.3 and various M_a } we see that for increasing $M_a$ the excitations converge towards the results obtained for a delta potential impurity. Already for ${M_a = 16}$ the behavior of the HA-like mode is sufficiently similar to the first excitation of the fixed delta potential impurity that it appears as though the bosons experience the finite mass impurity as a fixed potential and the continuous rotational symmetry has been broken. One qualitative difference that distinguishes the system with an impurity of any finite mass from the delta potential impurity is the presence of NG-like modes. These modes appear for any finite impurity mass but are absent in the case of a fixed delta potential impurity.
We interpret the apparent convergence of the HA-like modes towards the first excitation of an infinitely massive impurity system as a transition from a spontaneous to an explicitly broken rotational symmetry of the Hamiltonian. Furthermore, with increasing $M_a$ the characteristic minimum of the HA-like mode becomes less and less prominent, as we see a flattening of the mode for weak $g_{ab}$. However, despite this flattening, the excitation mode of the delta potential impurity system looks similar to the case of spontaneous symmetry breaking, with a negative slope for weak $-g_{ab}$, albeit a shallow one. The minima of the excitation mode for the fixed delta potential shown in Fig.~\ref{fig: first excitation for N = 6 g = 0.3 and various M_a } is due to the density of the bosons in the first excitation localizing and thereby increasing their overlap with the impurity potential as $-g_{ab}$ is increased.

We can understand the convergence behavior in the non-interacting impurity limit, ${g_{ab} = 0}$, by considering individual particle excitations. At least a two-particle excitation is required to conserve angular momentum.
For ${M_a > 1}$ and sufficiently weak interactions it is energetically favorable to excite the impurity and a single boson. For a non-interacting impurity the excitation energy of the system is simply ${\Delta E=E_\textrm{LL}+{1}/{(2M_a)}}$ where $E_\textrm{LL}$ is the excitation energy of the the repulsive bosons on a ring from the Lieb-Liniger model~\cite{PhysRev.130.1605}. As $M_a$ increases, the excitation energy gradually decreases, scaling as ${\propto {1}/{M_a}}$  and approaches the limit of the 
explicitly broken rotational symmetry,
as can be seen in Fig~\ref{fig: first excitation for N = 6 g = 0.3 and various M_a }.
For ${M_a < 1}$
and ${g_{ab} = 0}$ with sufficiently small $g_{bb}$, it is energetically favorable to excite two bosons. Therefore the impurity mass has no influence on the excitation energy in this limit, leading to the degeneracy that we see in Fig.~\ref{fig: first excitation for N = 6 g = 0.3 and various M_a } for the ${M_a < 1}$ modes at ${g_{ab} = 0}$. Note that in the completely non-interacting limit ${(g_{bb} = g_{ab} = 0)}$ we would also see this degeneracy for the $M_a = 1.0$ mode.

\begin{figure}
    \centering
    \includegraphics[width=\columnwidth]{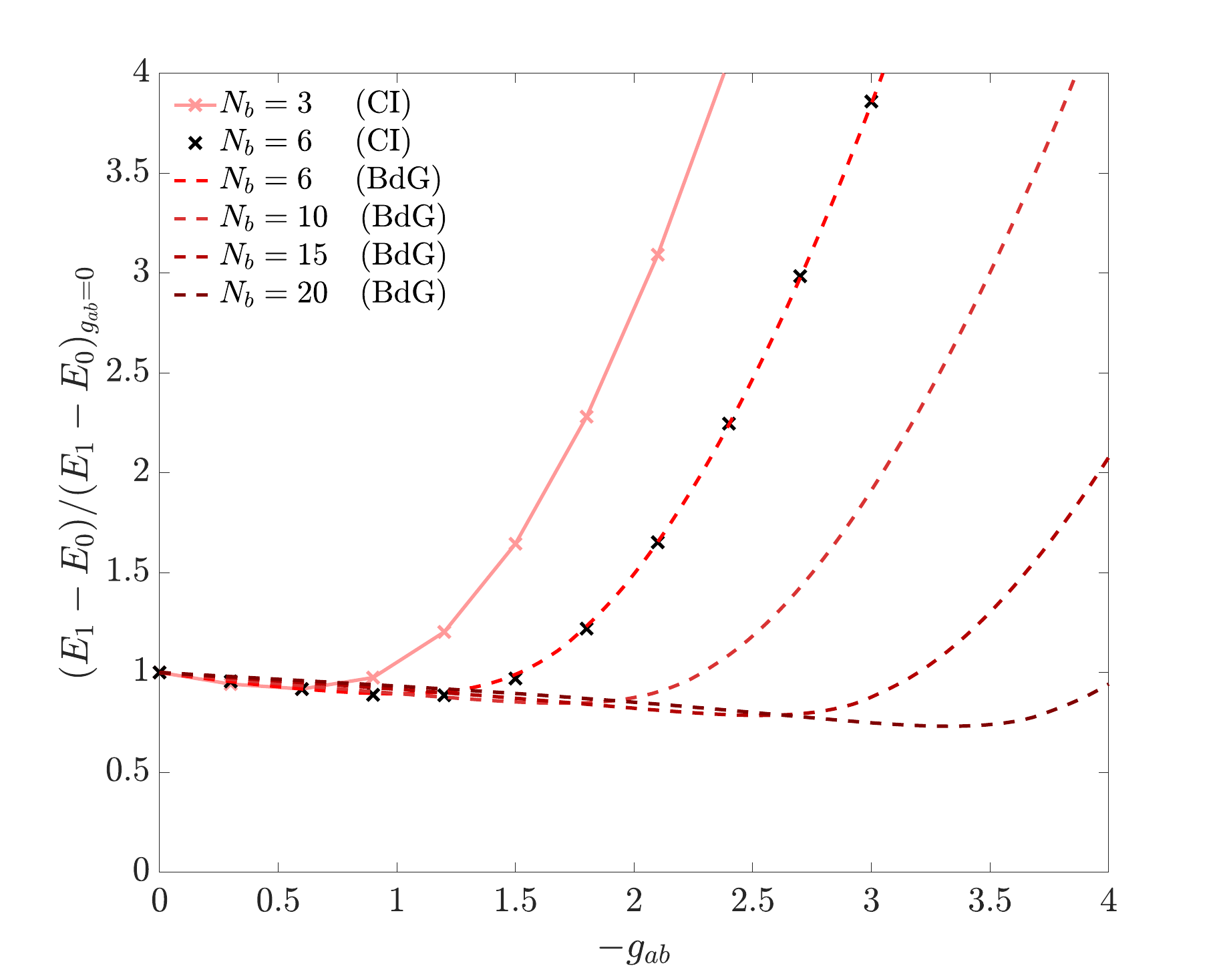}
    \caption{First excitation of repulsive $(g_{bb} = 0.3)$ bosonic systems with a delta potential impurity for various boson numbers, $N_b$, as a function of boson-impurity attraction. The excitations are normalized by their respective excitation energies in the non-interacting impurity limit $(g_{ab} = 0)$. The exact diagonalization results $(N_b = 3,6)$ are plotted as markers with connecting lines to guide the eye. The dashed lines $(N_b = 6,10,15,20)$ show the first BdG excitation of the numerical ground state solutions to the associated GPEs. The BdG excitations are computed in steps of $0.04$.}
    \label{fig: BdG modes with increasing N}
\end{figure}

Another effect of increasing $M_a$ is a shift of the minimum of the HA-like mode to lower values of $-g_{ab}^*$. This suggests that as $M_a$ increases, the bosons are more easily bound to the impurity. We interpret this as follows: For larger $M_a$ the kinetic energy contribution of the impurity is smaller, making the system effectively more attractive, or the relative contribution of the $g_{ab}$ term in Eq.~\eqref{eqn: hamiltonian} larger. This allows for more bosons to bind to the impurity with increasing $M_a$, contrary to what one expects from MF analysis where the critical condition is independent of the boson and impurity masses.
Furthermore, this analysis is specific to bosonic systems and in fact is in contrast to the case of fermions for which Ref.~\cite{PhysRevA.106.L011302} suggests that indeed lighter impurities bind more fermions. 

For ${g_{ab} < g_{ab}^*}$ the modes increase {quadratically} with $-g_{ab}$. We can understand this from the mean-field solution.
We assume that for $g_{ab} < g_{ab}^*$ all bosons are bound to the impurity, and in order to excite the system while conserving $L$ we must overcome the bound-state energy, which from mean-field results we know to be proportional to ${g_{ab}^2}$~\cite{Brauneis_2022}.

Figure~\ref{fig: BdG modes with increasing N} shows the first excitation for the repulsive ${(g_{bb} = 0.3)}$ bosonic system with an infinitely massive impurity as a function of impurity-boson attraction strength ${(-g_{ab})}$ for various $N_b$. The markers indicate the first excitation for the few-body systems $(N_b = 3,6)$ computed in the CI framework, with connecting lines to guide the eye. The dashed lines show the first BdG excitation of the numerical ground state solution to the GPE. The agreement between the few-body and MF results is noteworthy. For as few as ${N_b=6}$ bosons the infinitely massive impurity system is well described by the MF BdG excitations, showing good agreement with the much more computationally costly CI approach.

\subsection{Rotational Spectra}

\begin{figure}[t!]
    \centering
    \includegraphics[width=0.95\columnwidth]{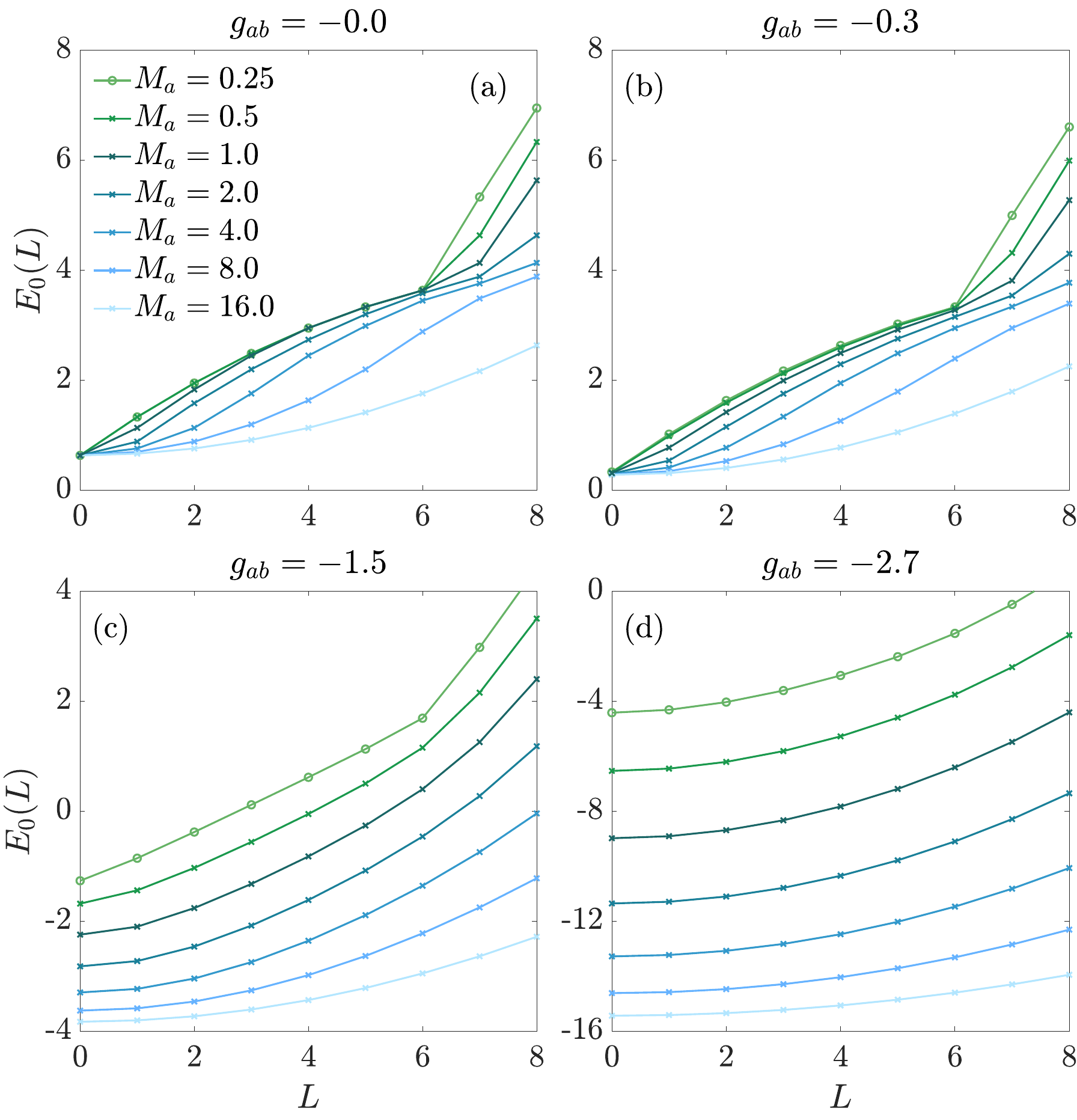}
    \caption{Ground state energy as a function of total angular momentum, $L$, for a system of ${N_b = 6}$, ${g_{bb} = 0.3}$,  and various impurity masses, ${0.25 \leq {M_a} \leq 16.0}$ with boson-impurity attraction ${g_{ab} = 0.0}$ (a), $-0.3$ (b), $-1.5$ (c), $-2.7$ (d). } 
    \label{fig: yrast Lines for N = 6 g = 0.3 and various M_a }
    \centering
\end{figure}

Let us now consider the effect of the impurity mass on the rotational spectrum. Figure~\ref{fig: yrast Lines for N = 6 g = 0.3 and various M_a } presents the ground state energies as a function of the total angular momentum (the so-called ``yrast" lines) for a system of ${N_b = 6}$ bosons with ${g_{bb} = 0.3}$, ${g_{ab} = 0.0, -0.3, -1.5,}$ and $-2.7$, and various impurity masses ${0.25 \leq M_a \leq 16.0}$. 
The effect of $M_a$ on the yrast line curvature is most notable in the limit of a weakly attractive impurity.
Intuitively, this makes sense, as for sufficiently large $-g_{ab}$ we expect the system to be bound and thus for the yrast lines to exhibit the parabolic curvature associated with rigid body rotation for all considered values of $M_a$.  
In Fig.~\ref{fig: yrast Lines for N = 6 g = 0.3 and various M_a }~(d) where ${-g_{ab} = 2.7 > -g_{ab}^*}$ for all $M_a$ considered, we see this parabolic curvature of the yrast lines. 
By increasing $M_a$ the contribution of the impurity to the kinetic energy, the first term of Eq.~\eqref{eqn: hamiltonian}, is reduced and thus the relative contribution of the attractive boson-impurity interaction, the final term of Eq.~\eqref{eqn: hamiltonian} becomes more significant, driving the system to localize and causing the ground state energy (the offset of the parabolic yrast lines) to decrease with increasing $M_a$. 
Furthermore, we expect the first derivative of each parabolic yrast line to be inversely proportional to the total mass of the system, ${\propto (M_a + N_bM_b)^{-1}}$, resulting in the flattening of the yrast lines with increasing $M_a$.

Let us next consider the case of the non-interacting impurity $(g_{ab} = 0)$ presented in Fig.~\ref{fig: yrast Lines for N = 6 g = 0.3 and various M_a }~(a). A notable feature of this figure is the degeneracy of all yrast lines at ${L = 0}$.
For $L = 0$, a non-interacting impurity will occupy the zero angular momentum one-body state, $\phi_0$, and therefore not contribute to the energy of the system, leading to a degeneracy for all $M_a$. 
For ${M_a < M_b}$ this degeneracy of the $g_{ab} = 0$ yrast lines persist for all ${L \leq N_b}$. This behavior
is best explained by considering single particle excitations. 
The one-body basis states $\phi_m$ given by Eq.~\eqref{eqn: one-body basis} have associated one-body energy $e_m = {m^2}/{(2M_\sigma)}$ with ${\sigma\in\{a,b\}}$.
Clearly, for a single-particle excitation it is always energetically favorable to excite a heavier particle. For ${L > 1}$, due to the one-body excitation energies being quadratic in angular momentum, for sufficiently weak interactions it is energetically favorable to excite multiple particles to $\phi_1$, rather than excite a single particle to a higher one-body angular momentum state. Therefore, when the impurity mass is less than the boson mass ${(M_a < 1.0)}$ for a non-interacting impurity and sufficiently weak boson-boson repulsion, the system favors the excitation of multiple bosons to the ${m=1}$ one-body basis state while the impurity remains in $\phi_0$. 
The result is the degeneracy of all yrast lines for $M_a < M_b$ and $L \leq N_b$ when the impurity is non-interacting ${(g_{ab} = 0)}$.
For ${L > N_b}$ it becomes energetically favorable to excite all bosons and the impurity to single-particle orbitals with $m > 0$. Thus, the yrast lines will not be degenerate for these values of $L$ for any unequal $M_a$.
For ${M_a > M_b}$ excitations of the impurity become energetically favourable for any ${L > 0}$, breaking the degeneracy and shifting the inflection of the yrast line curvature to higher $L$ with increasing $M_a$.

Figure~\ref{fig: yrast Lines for N = 6 g = 0.3 and various M_a }~(b) includes weak boson-impurity attraction, ${g_{ab} = -0.3}$. In this case there is no degeneracy of the yrast lines at $L = 0$, but the energy differences are smaller than the marker size and therefore not resolved in the figure. Similarly, the degeneracy of the yrast lines with ${M_a < M_b}$ for ${L \leq N_b}$ is also lifted, though the energy difference is not resolved in the figure.

For all values of $g_{ab}$, Fig.~\ref{fig: yrast Lines for N = 6 g = 0.3 and various M_a }~(a)-(d), the case ${M_a = M_b}$ has some unique symmetrical properties.
As in the case of a balanced two-component system, due to the periodic boundary conditions of the ring, for equal impurity and boson masses it follows from Bloch's theorem~\cite{Bloch_PhysRevA.7.2187} that the rotational spectrum is composed of a component that is periodic in ${L}$ with periodicity ${N = N_b + 1}$ and a component that is quadratic in $L$. Furthermore, the periodic component of the spectrum is symmetric about $N/2$~\cite{Bloch_PhysRevA.7.2187,PhysRevLett.103.100404}. These features of the energy spectra can be shown using the fact that total angular momentum, $L$, is a conserved quantity and by considering the transformations $m\rightarrow m + r$ for arbitrary integer $r$ and $m \rightarrow 1 - m$ of all one-body angular momenta, $m$, for all many-body basis states with a given total angular momentum, $L$. By considering these transformations one can also see how the periodicity and symmetry of the spectra breaks down for ${M_a \neq M_b}$. 
In particular, the transformation ${m\rightarrow 1 - m}$ of all one-body angular momentum quantum numbers maps many-body basis sates with total angular momentum $L$ and energy $E$ to states with total angular momentum $L' = N - L$ and energy
\begin{equation}\label{eqn: E' = E - ...}
    E' = E - \frac{L}{M_b} + \frac{1}{2}\Big(\frac{1}{M_a} + \frac{N_b}{M_b}\Big) + \Big( \frac{1}{M_b} - \frac{1}{M_a} \Big)m_a.
\end{equation}
Here, $m_a$ is the one-body angular momentum eigenvalue of the impurity in the original many-body basis state of energy $E$ and total angular momentum $L$. In the equal mass case, in units of ${M_b=1}$, we recover ${E' = E - L + N/2}$ from which it is seen that the periodic component of the rotational spectrum is symmetric about $N/2$~\cite{Bloch_PhysRevA.7.2187,PhysRevLett.103.100404}. For $M_a\neq M_b$ the final term of Eq.~\eqref{eqn: E' = E - ...} depends on the details of the initial state. We no longer have a uniform shift in energy of all many-body basis states in the subspace of the Hilbert space defined by total angular momentum $L$. Thus the energy spectrum no longer has a component that is symmetric about $L = N/2$ and the energy term responsible for the deviation from the symmetry is proportional to $(\tfrac{1}{M_b} - \tfrac{1}{M_a})$.
In a similar way, it can be seen from the transformation ${m\rightarrow m+r}$ that the periodic component of the energy spectrum depends on the equality of the boson and impurity masses. Furthermore, for ${M_a \neq M_b}$ the energy term that destroys the periodicity of the spectra is also proportional to $(\tfrac{1}{M_b} - \tfrac{1}{M_a})$.

\section{Summary and Outlook}\label{Sec: Summary}

In summary, we have seen that the presence of a single attractive impurity can drive localization in a repulsive few-body bosonic system. This localization is observed in the density of the bosons in the co-moving frame of the impurity and accompanied by signatures of finite-size precursors of HA-like and NG-like modes in the low-lying energy spectrum. 
We have found that with increasing impurity mass the behavior of the HA-like mode for the finite impurity mass system converges towards the behavior of the first excitation of a system of repulsive bosons in the presence of an infinitely massive impurity, or equivalently, a fixed attractive delta potential. We interpret this convergence behavior as a transition from a spontaneous to an explicit breaking of the continuous rotational symmetry of the system Hamiltonian.
In the mean-field limit the BdG excitations of the ground state solution to the GPE of a repulsive bosonic system with a fixed delta potential impurity show remarkable agreement with the few-body CI calculations for as few as ${N_b = 6}$ bosons.
The effect of increasing impurity mass on driving the localization was further demonstrated by the analysis of the rotational spectra of the few-body system.

From the exact few-body spectra of the impurity system we have identified the minima of the HA-like mode as a marker of the onset of localization. 
The condition Eq.~\eqref{eqn: expectation of Hamiltonian derivative} is independent of the form of interaction driving the transition, while
Eq.~\eqref{eqn: equality of pair correlations} is applicable to systems where the phase transition or few-body phase crossover may be driven by a delta-potential interaction, 
and may be useful in the future study of systems such as the pairing of fermions in the few-to many-body regime~\cite{Bjerlin2016, PhysRevA.110.L061302, Bayha2020},
soliton formation in a BEC~\cite{kanamoto2003quantum, Ueda_HA_GS_PhysRevLett.94.090404}, the BEC to supersolid phase transition~\cite{hertkorn2019fate} or binary bosonic systems supporting quantum droplet formation~\cite{Chergui_2023}.

\begin{acknowledgments}
We thank K. Mukherjee and P. St\"{u}rmer for valuable discussions and input on the numerical approach to calculating the Bogoliubov de Gennes excitations. We thank J. Bengtsson for his work on the exact diagonalization library. We thank Hans-Werner Hammer for helpful discussions. 
This research was financially supported by the Knut and Alice Wallenberg Foundation (Grant No. KAW 2018.0217), the Swedish Research Council (Grant No. 2022-03654\_VR), a fellowship of the German Academic Exchange Service (DAAD) and by NanoLund. 
Part of the computations were enabled by resources provided by the National Academic Infrastructure for Supercomputing in Sweden (NAISS), partially funded by the Swedish Research Council through Grant Agreement No. 2022-06725\_VR.
\end{acknowledgments}

\appendix

\section{Delta Potential Impurity}\label{appendix: Delta Potential Trap}

In the infinitely massive limit $(M_a\rightarrow\infty)$, the impurity acts as a fixed delta potential on the ring, which explicitly breaks the continuous rotational symmetry of the Hamiltonian. 
It can be approximated as an infinitesimally narrow Gaussian (see {e.g.}~\cite{arfken2013mathematical})
\begin{equation}
     \delta (\theta_b) = \lim_{\alpha\xrightarrow[]{}\infty}\sqrt{\frac{\alpha}{\pi}}\exp(-\alpha(\theta_b)^2).
\end{equation}
Without loss of generality we have taken the position of the delta potential to be ${\theta_a = 0}$.
To compute the energy eigenstates in this limit the kinetic term of the impurity is neglected and we take the dimensionless paramter ${\alpha = 10^7}$. We no longer use the angular momentum eigenstates and instead take the one-body basis states to be the energy eigenstates of a single boson in the presence of the infinitely massive impurity, 
\begin{equation}
    \hat{h}(\theta_{b,i}) = -\frac{1}{2}\frac{\partial^2}{\partial \theta_{b,i}^2} + g_{ab}\delta(\theta_{b,i}).
\end{equation}
Here $g_{ab}$ has been renormalized in connection with the finite basis size as described in Ref.~\cite{brauneis2024}. 
These one-body basis states are constructed from 503 B-splines of seventh order~\cite{de1978practical}.  503 B-splines were used to accommodate a non-linear distribution of breakpoints around the ring. The largest concentration being in the vicinity of the impurity.
Half of the breakpoints are distributed linearly within ${|\theta| \leq \frac{\pi}{4}}$, one quarter of the breakpoints are distributed linearly within ${\frac{\pi}{4} < |\theta| \leq \frac{\pi}{2}}$ and the final quarter within ${\frac{\pi}{2} < |\theta| \leq \pi}$.
The one-body basis is truncated to include the 30 lowest one-body energy eigenstates from which the many-body states are constructed.
The many-body Hamiltonian, Eq.~\eqref{eqn: hamiltonian} (excluding the first term and setting $\theta_a=0$) is diagonalized in the resulting Hilbert space. 
To demonstrate the convergence obtained with this procedure, the ground state energy and excitation spectra are shown as a function of the number of one-body basis states $n_\text{OBB}$ in Fig.~\ref{fig: Convergence N = 6, M_a = Delta, g = 0.3, gab = -3.3} for a system of $N_b = 6$ repulsive $(g_{bb} = 0.3)$ bosons in the presence of a fixed attractive $(g_{ab} = -3.6)$ delta potential.
\begin{figure}[t!]
    \centering
    \includegraphics[width=0.45\textwidth]{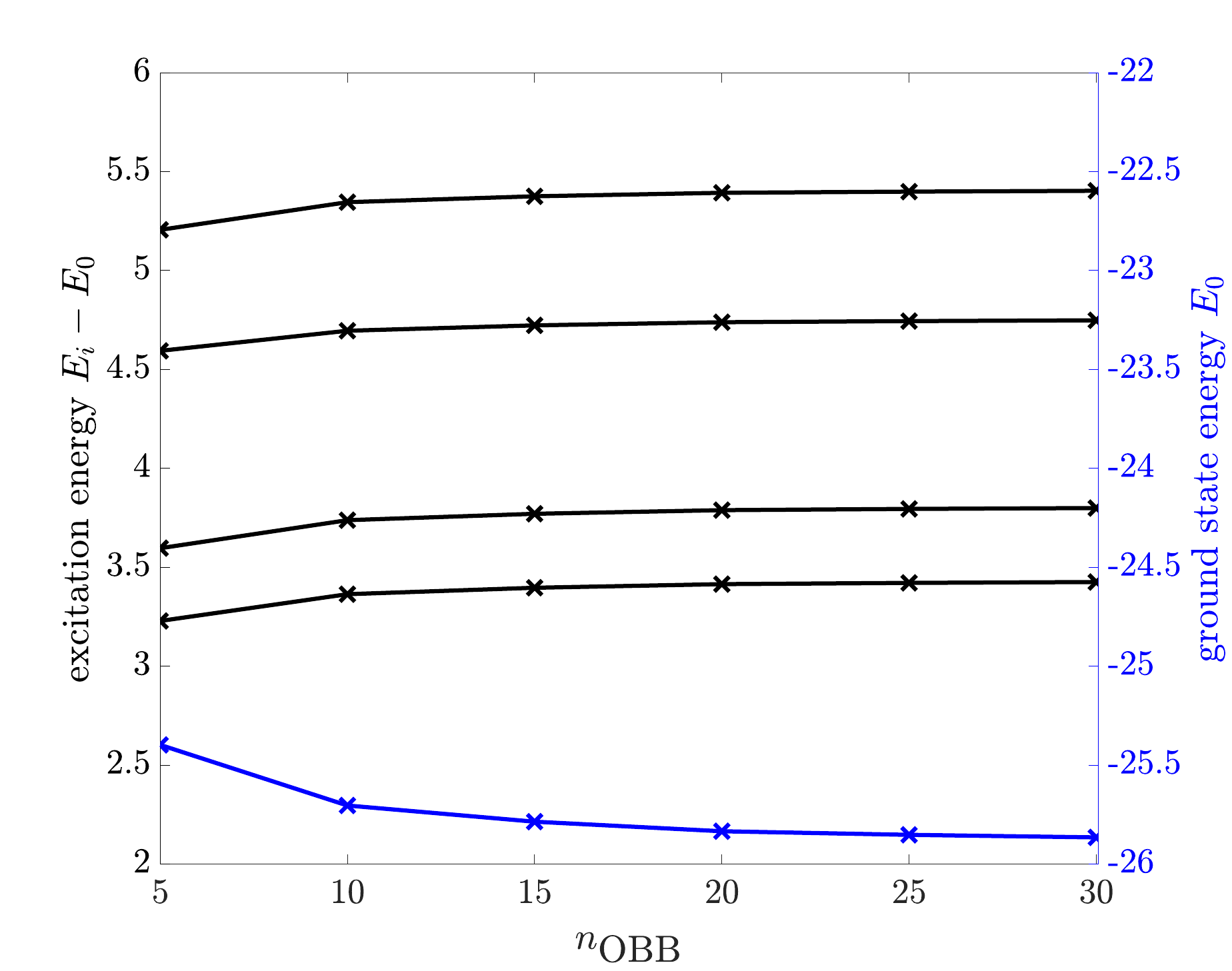}
    \caption{Excitation energy spectra (black) and ground state energy (blue) as a function of one-body basis size, ${n_\textrm{OBB}}$, for a system of ${N_b = 6}$ repulsive ${(g_{bb} = 0.3)}$ bosons and a fixed attractive ${(g_{ab} = -3.6)}$ delta potential. These figures indicate the convergence of the delta potential results in Fig.~\ref{fig: first excitation for N = 6 g = 0.3 and various M_a }.} 
    \label{fig: Convergence N = 6, M_a = Delta, g = 0.3, gab = -3.3}
    \centering
\end{figure}

\section{Convergence}\label{Appendix: Convergence}

In the main text, the energy spectra and pair correlations for finite mass impurity systems are computed via exact diagonalization/configuration interaction (CI) methods. 
The one-body basis is taken to be the angular momentum eigenstates $\phi_m(\theta) = \frac{1}{\sqrt{2\pi}}e^{im\theta}$ with truncation $|m| \leq m_\textrm{Max} = 60$ for integer one-body angular momentum $m$. The resulting many-body basis states are the total angular momentum eigenstates which block diagonalize the system Hamiltonian. Thus the Hamiltonian is diagonalized for each value of total angular momentum separately. A so-called importance-truncation configuration interaction (ITCI) scheme~\cite{PhysRevC.79.064324,Tubman2020} similar to that employed  in Ref.~\cite{Chergui_2023} is utilized. (In the present work we account for unequal particle numbers and unequal masses in the two components.) A subspace of the total Hilbert space tailored to the target eigestate is iteratively updated until a desired level of convergence is achieved.  Following the procedure of Ref.~\cite{PhysRevC.79.064324} the total Hilbert space $\mathcal{H}$ is partitioned into a reference subspace $\mathcal{H}_\text{ref}$ and the orthogonal complementary subspace $\mathcal{H}_C$. 
\begin{figure}[h!]
    \centering
    \includegraphics[width=0.45\textwidth]{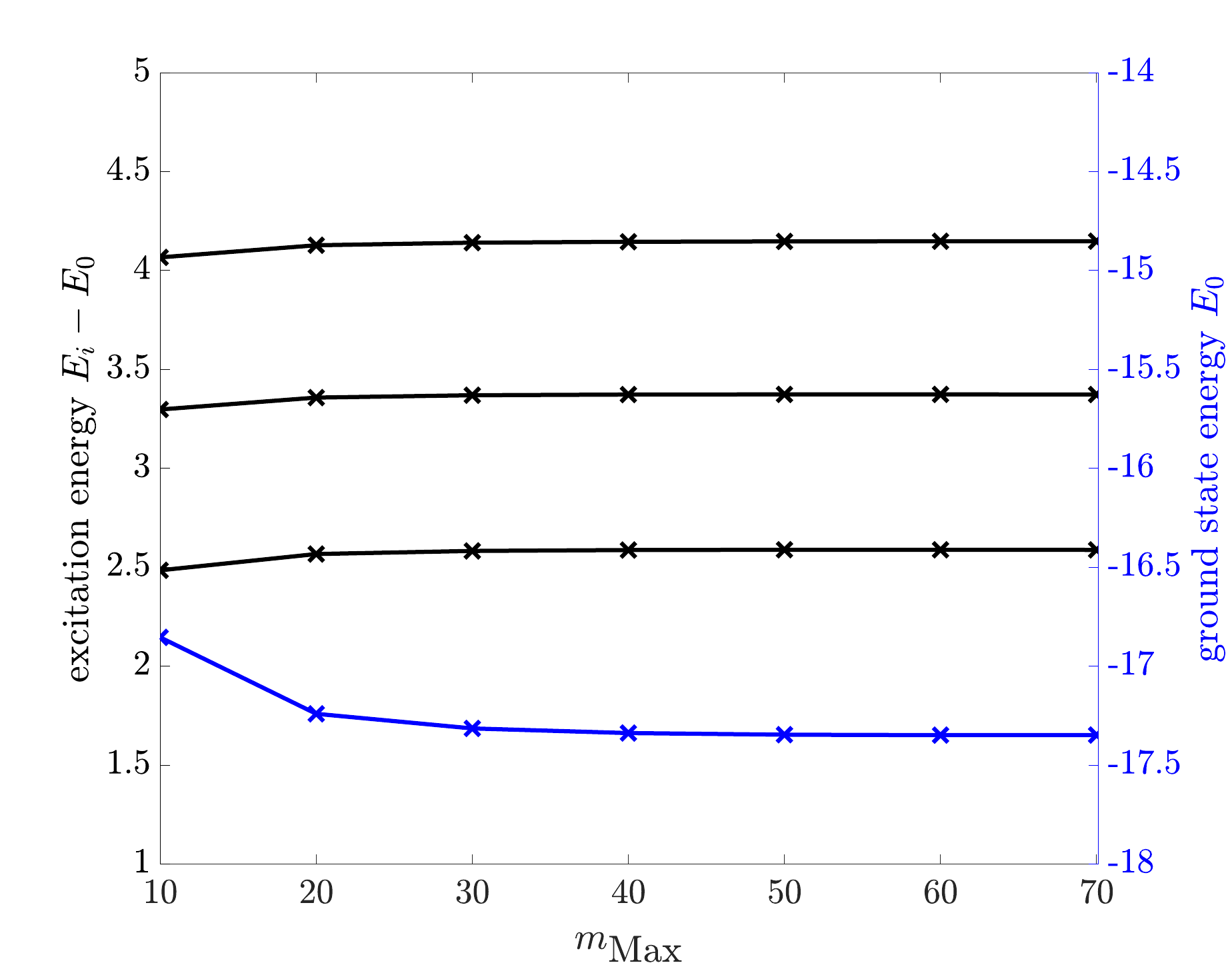}
    \caption{Excitaiton energy spectrum (black) and ground state energy (blue) as a function of one-body angular momentum cutoff, ${m_\textrm{Max}}$, for a system of $N_b = 6$ bosons and an impurity of mass ${M_a = 1.0}$ with boson-boson repulsion ${g_{bb} = 0.3}$ and boson-impurity attraction ${g_{ab} = -3.6}$. This figure indicates the convergence of Figs.~\ref{fig: pair correlations N = 6 g = 0.3},~\ref{fig: first excitation for N = 6 g = 0.3 and various M_a },~\ref{fig: yrast Lines for N = 6 g = 0.3 and various M_a }.}
    \label{fig: convergence N = 6 g = 0.3 and Ma = 1 }
\end{figure}
\begin{figure}
    \centering
    \includegraphics[width=0.45\textwidth]{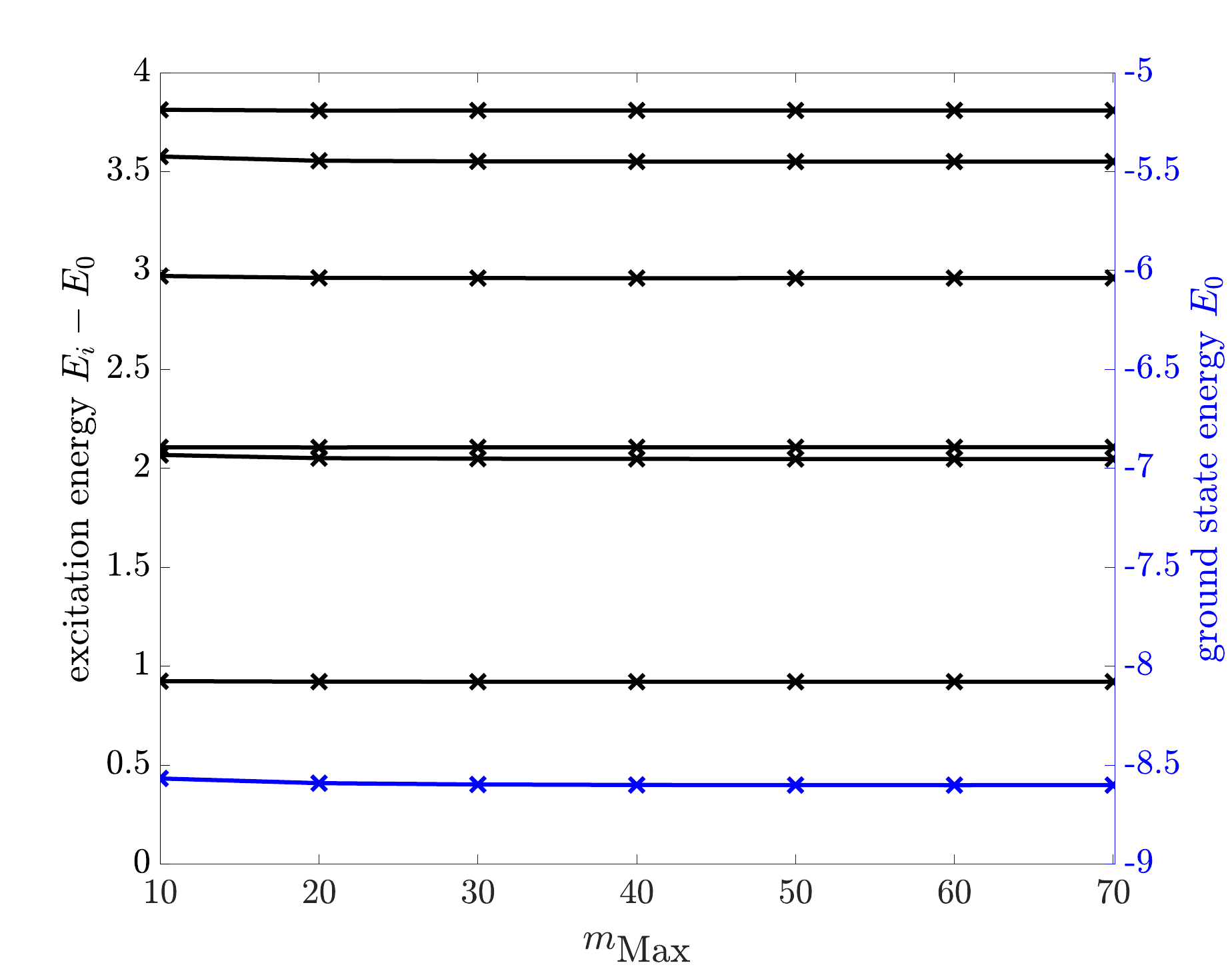}
    \includegraphics[width=0.45\textwidth]{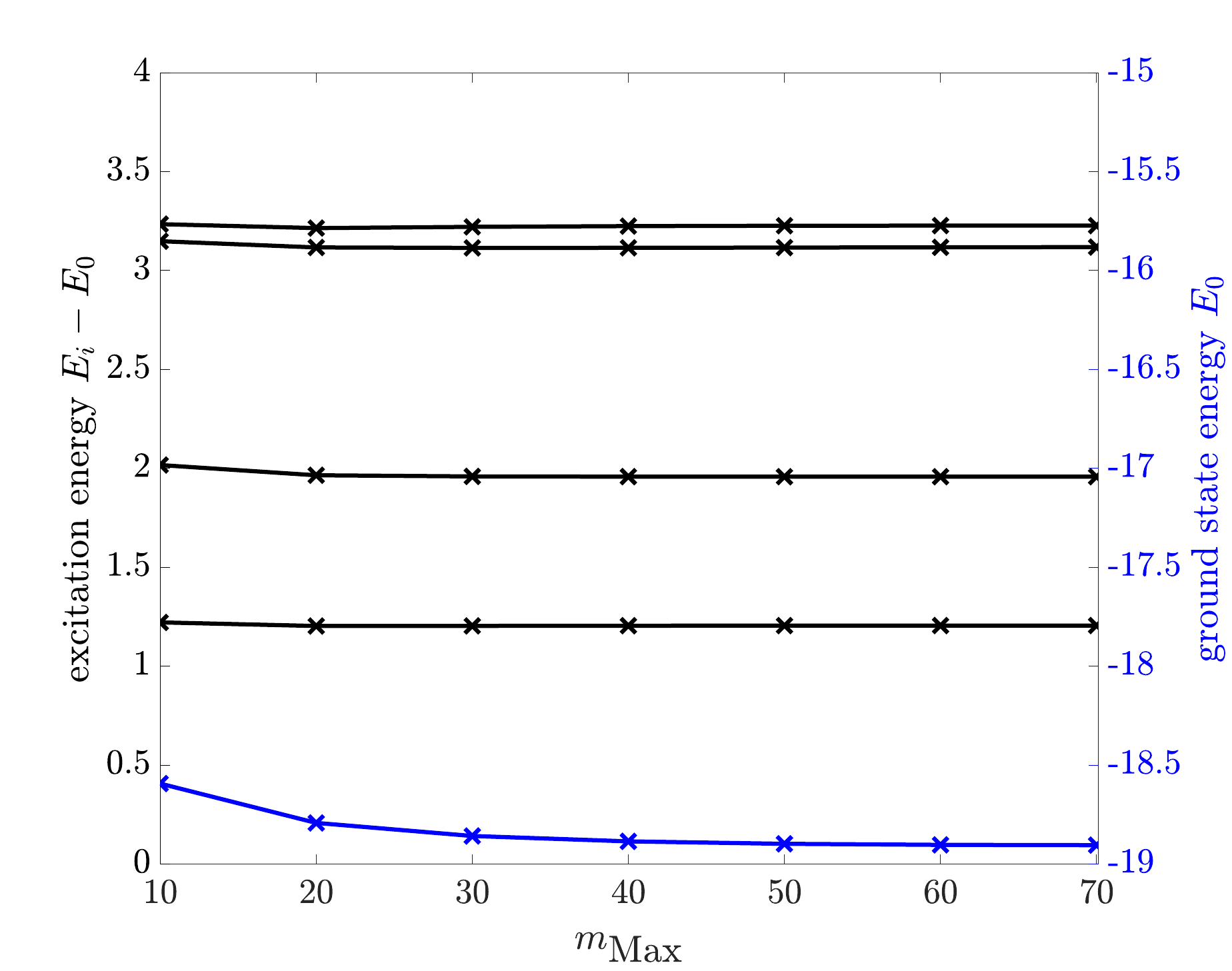}
    \includegraphics[width=0.45\textwidth]{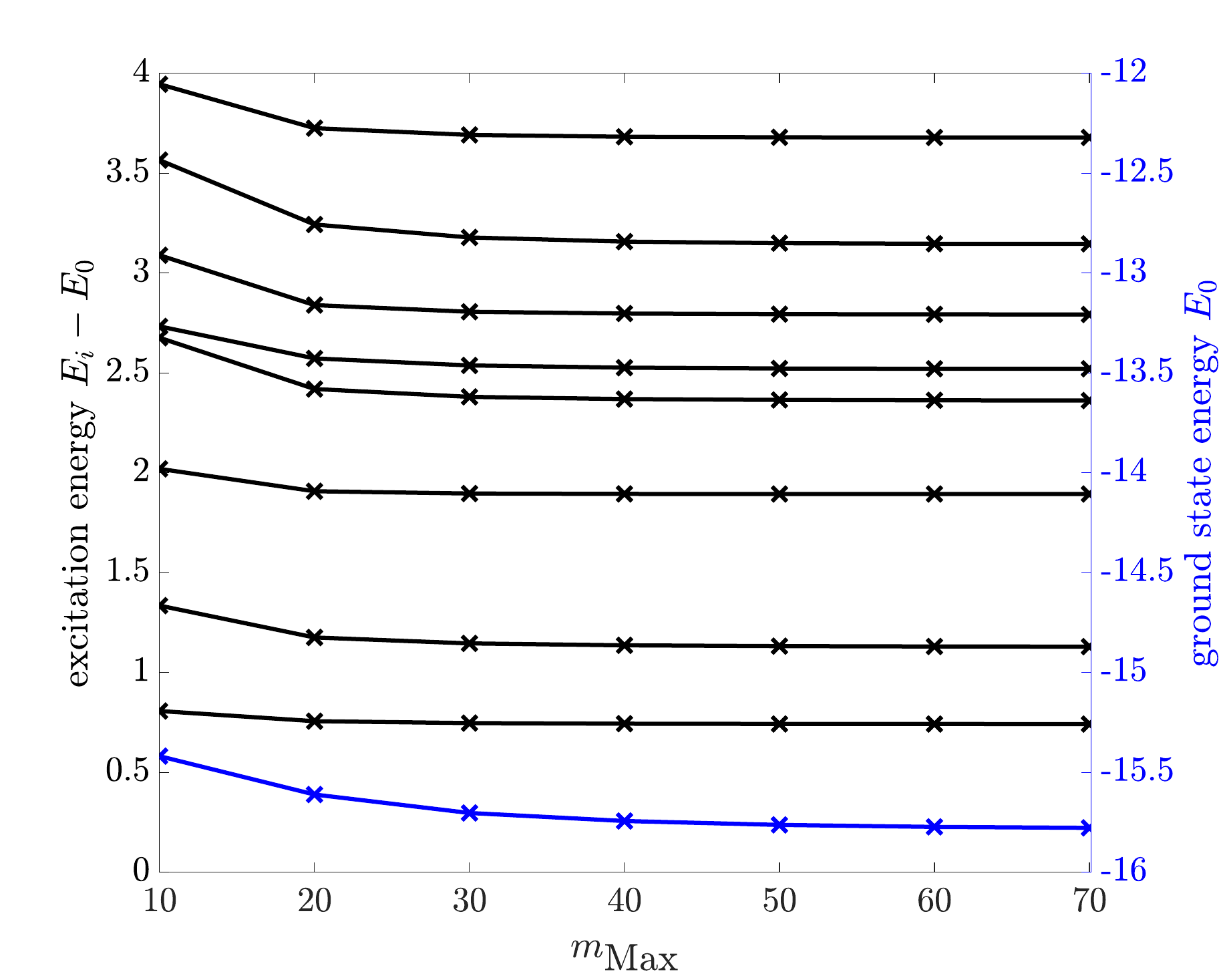}
    \caption{Excitation energy spectrum (black) and ground state energy (blue) as a function of one-body angular momentum cutoff, ${m_\textrm{Max}}$, for a system of $N_b = 6$ bosons and an impurity of mass ${M_a = 0.25, 1.0, 16.0}$ (top, middle, bottom) with boson-boson repulsion ${g_{bb} = 1.2}$ and boson-impurity attraction ${g_{ab} = -4.8}$ (top, middle row) and ${g_{ab} = -3.6}$ (bottom row). These figures indicate the convergence of Fig.~\ref{fig: MF Comparison}.}
    \label{fig: convergence N = 6, Ma = 0.25, 1.0, 16.0, with gbb = 1.2 and gab = -4.8, -4.8, and -3.6 respectively}
\end{figure}
The Hamiltonian is diagonalized in $\mathcal{H}_\text{ref}$ and the target eigenstate wave function is expanded into $\mathcal{H}_C$ using  multiconfigrational first order perterbation theory with Epstein-Nesbet-like partitioning~\cite{Epstein1926,Nesbet1955}.
As in Ref.~\cite{Chergui_2023}, the importance measure, $\kappa_\nu$, of a given element of $\mathcal{H}_C$ is taken to be its dimensionless perturbative amplitude in the expansion of the target eigenstate into $\mathcal{H}_C$. The importance threshold for transferring a state from $\mathcal{H}_C$ to $\mathcal{H}_\text{ref}$ is $\kappa_\nu > \kappa_\text{Min} = 10^{-5}$. Only states in $\mathcal{H}_\text{ref}$ with amplitude greater that $C_\textrm{Min} = 10^{-4}$ in the target eigenstate are used when computing importance measures. The condition to terminate the iterative search for relevant basis states is that the relative difference between the energy of the target eigenstate between two iterations is less than $10^{-5}.$
{Figures~\ref{fig: convergence N = 6 g = 0.3 and Ma = 1 }-\ref{fig: convergence N = 6, Ma = 0.25, 16.0, gbb = 0.3 and gab = -3.6} demonstrate the level of convergence obtained with the ITCI approach for the systems considered in the main text. They show the ground state energy and low-lying excitation spectra for various impurity masses and and interaction strengths as a function of the one-body basis truncation $m_\textrm{Max}$.}

In addition to the ITCI method we utilize running coupling constants to further aid in the convergence of the exact diagonalization results. We employ the method of renormalization of the contact interaction described in Ref.~\cite{brauneis2024} for the illustrative system of fermions in a one-dimensional harmonic trap. Here we apply the same procedure to the contact interactions between bosons on a one-dimensional ring. The main idea is to consider the interaction parameters of the Hamiltonian as being dependent on the choice of one-body angular momentum cutoff $m_\textrm{Max}$ (see Sec.~\ref{Sec: Model}). The ground state energy of two bosons interacting on a one-dimensional ring has an exact analytical solution corresponding to ${m_\textrm{Max}\rightarrow\infty}$. For each interaction parameter, $g_{\sigma\sigma'}$, of the Hamiltonian, Eq.~\eqref{eqn: hamiltonian}, of the main text, we first find the exact analytical ground state energy of two bosons interacting via contact interactions with strength $g_{\sigma\sigma'}$ on the one-dimensional ring. Next, we find the value of the parameter $\Tilde{g}_{\sigma\sigma'}$ that reproduces the analytical two-body ground state energy to desired precision in the finite Hilbert space with $m_\textrm{Max} = 60$. This is done with a binary search in a window around the original interaction parameter $g_{\sigma\sigma'}$ where the two-boson Hamiltonian is diagonalized in the Hilbert space with cutoff $m_\textrm{Max}$ at each step of the binary search until a value for $\Tilde{g_{\sigma\sigma'}}$ is found that reproduces the analytical two-body energy to within a relative energy difference of $10^{-12}$.
The ITCI calculations are then performed with the renormalized interaction parameters $\Tilde{g}_{\sigma\sigma'}$ and the one-body angular momentum cutoff $m_\textrm{Max} = 60$.

\begin{figure}[h]
    \centering
    \includegraphics[width=0.45\textwidth]{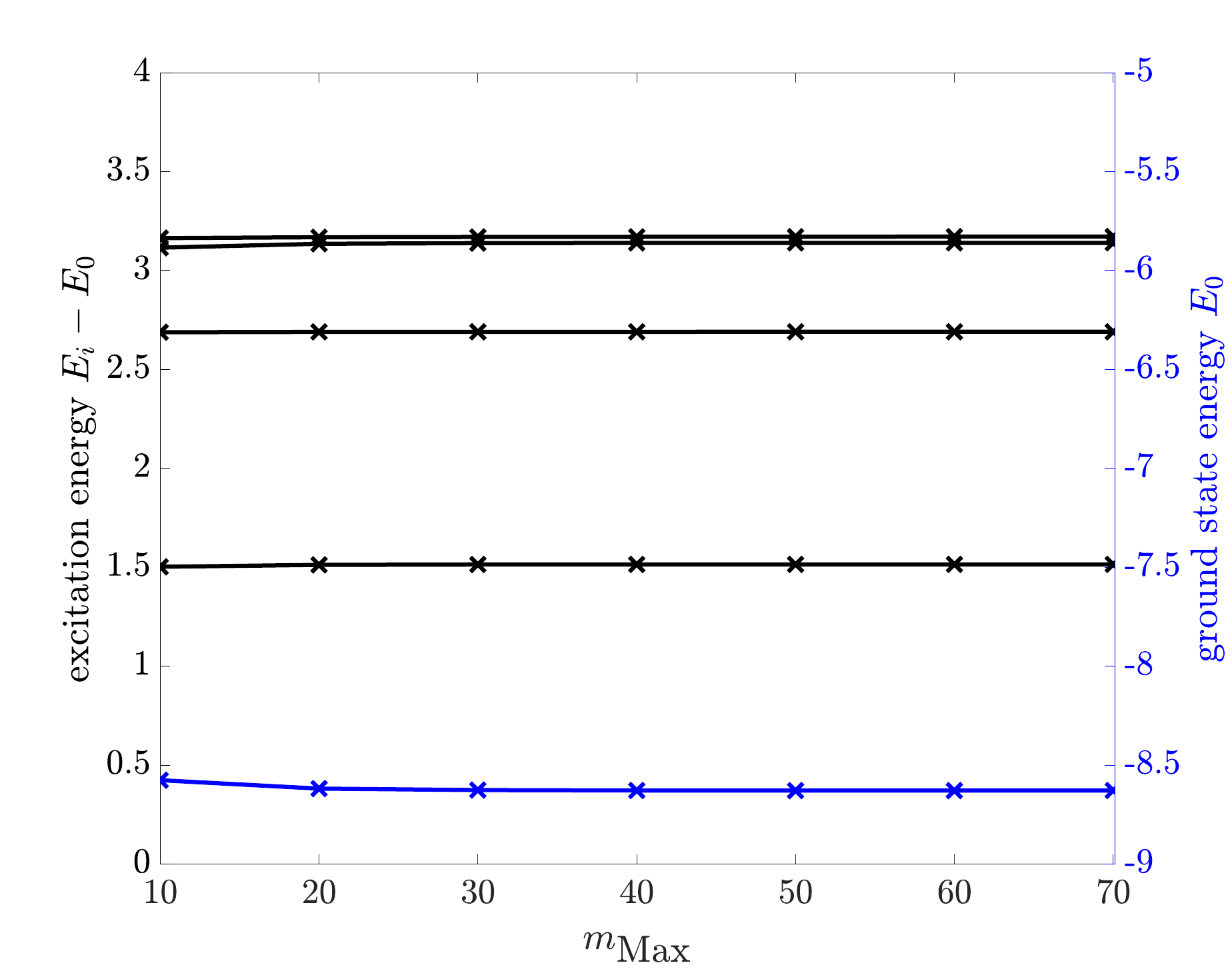}
    \includegraphics[width=0.45\textwidth]{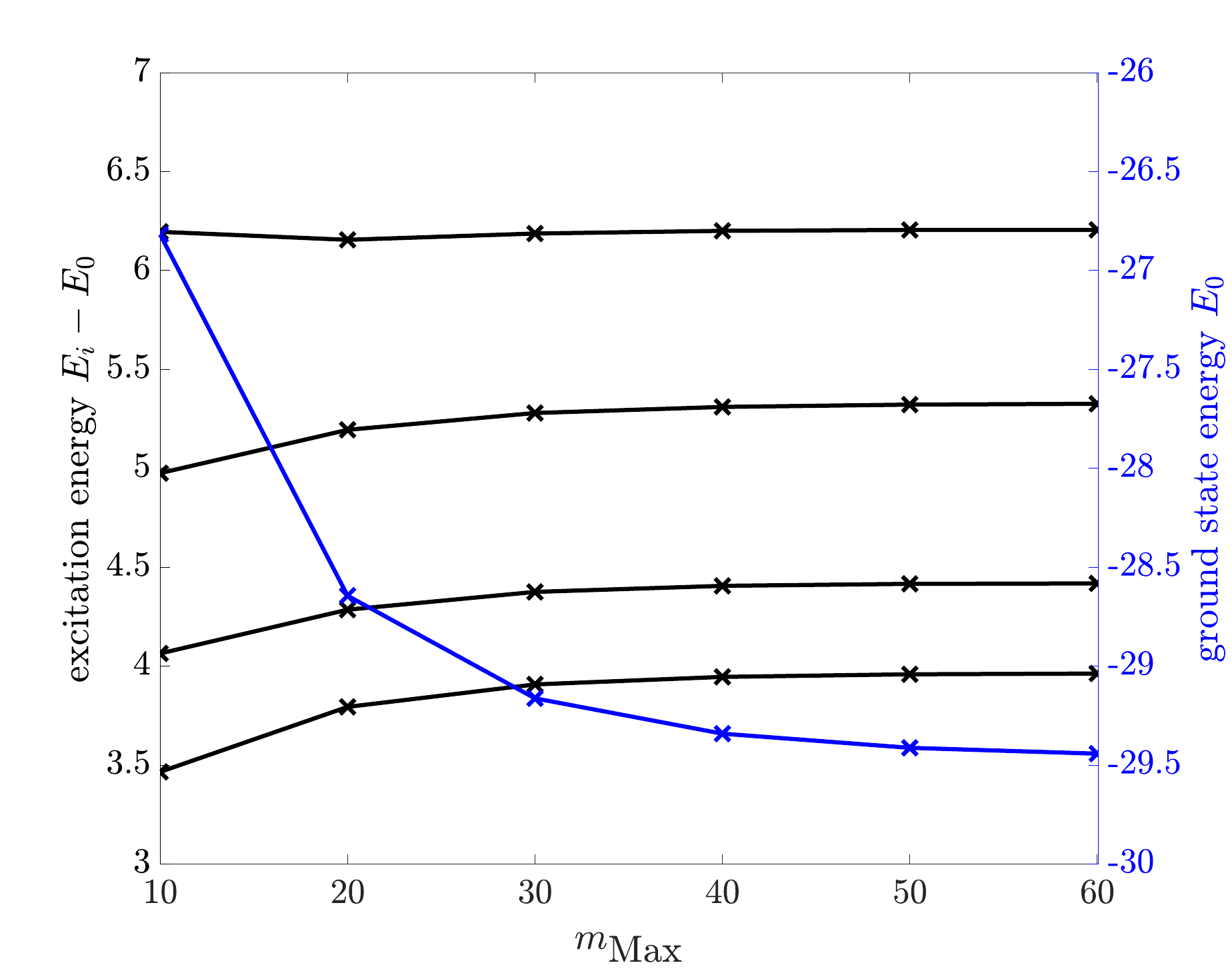}
    \caption{Excitation energy spectrum (black) and ground state energy (blue) as a function of one-body angular momentum cutoff, ${m_\textrm{Max}}$, for a system of ${N_b = 6}$ bosons and an impurity of mass ${M_a = 0.25, 16.0}$ (top, bottom) with boson-boson repulsion ${g_{bb} = 0.3}$ and boson-impurity attraction ${g_{ab} = -3.6}$. These figures indicate the convergence of Figs.~\ref{fig: first excitation for N = 6 g = 0.3 and various M_a },~\ref{fig: yrast Lines for N = 6 g = 0.3 and various M_a }.}
    \label{fig: convergence N = 6, Ma = 0.25, 16.0, gbb = 0.3 and gab = -3.6}
\end{figure}

\section{Determining the onset of localization from the Hellmann-Feynman theorem}\label{Appendix - PC Hellmann-Feynman}

In Section~\ref{Sec: pair correlations and hellmann-feynman} we argue that the minimum of the first excitation mode in the zero angular momentum subspace, ${L = 0}$, is a marker of the onset of localization. In this Appendix we show that this condition is equivalent to the equality of the pair correlations of the bosons in the ground state and first excited state at the fixed position of the impurity.
For convenience we here write the Hamiltonian in units defined by ${\hbar = M_b = R = 1}$,
\begin{equation}
    \begin{split}
       \hat{H} = -\frac{1}{2 M_a}\frac{\partial^2}{\partial\theta_{a}^2} &- \frac{1}{2}\sum_{i=1}^{N_b} \frac{\partial^2}{\partial \theta_{b,i}^2} + g_{bb}\sum_{i>j}\delta(\theta_{b,i} - \theta_{b,j})\\
       &+ g_{ab}\sum_{i=1}^{N_b}\delta(\theta_{b,i} - \theta_a ).
    \end{split} 
     \label{eqn: appendix hamiltonian}
\end{equation}
Here $M_a$ is the impurity mass, ${g_{bb} > 0}$ and ${g_{ab} < 0}$ are the boson-boson repulsion and impurity-boson attraction respectively and $\theta_a$ $(\theta_{b,i})$ is the azimuthal position of the impurity ($i^\textrm{th}$ boson) on the one-dimensional ring.
The pair correlations of the bosons with respect to the impurity at fixed position ${\theta'}$ for the system in state ${|\psi_i\rangle}$ is given by
\begin{equation}
    \begin{split}
        \rho^{(2,i)}_{ba}(\theta,\theta') = \sum_{m,n,k,l}\phi_m^*&(\theta)\phi_n^*(\theta')\phi_{k}(\theta')\phi_{l}(\theta) \\
        &\times\langle \psi_i| \hat{b}^\dagger_{m}\hat{a}^\dagger_{n}\hat{a}_{k}\hat{b}_{l}|\psi_i\rangle,
    \end{split}
    \label{eqn: appendix pair correlations}
\end{equation}
where $\hat{a}_m$ and $\hat{a}^\dagger_m$ ($\hat{b}_m$ and $\hat{b}^\dagger_m$) are the creation and annihilation operators for the impurity (boson) in the single-particle angular momentum eigenstate
with one-body angular momentum $m$.

The minima of the first excitation mode in the zero angular momentum subspace occurs when the first derivative of the ground state and first excited state energies are equal,
\begin{equation}
    \frac{\partial E_0}{\partial g_{ab}} = \frac{\partial E_1}{\partial g_{ab}}.
\end{equation}
From this fact, together with the Hellmann-Feynmann theorem~\cite{Hell-Feynman-PhysRev.56.340}
\begin{equation}
    \frac{dE^{(\lambda)}}{d\lambda} = \langle \psi^{(\lambda)} |\frac{d\hat{H}^{(\lambda)}}{d\lambda}|\psi^{(\lambda)}\rangle
    \label{eqn: appendix Hell-Feynman}
\end{equation}
were $E^{(\lambda)}$ and $\psi^{(\lambda)}$ are an eigenvalue and the corresponding eigenstate of the Hamiltonian $\hat{H}^{(\lambda)}$ and $\lambda$ is a parameter of the Hamiltonian, it follows that
\begin{equation}
    \langle{\psi_0|\frac{\partial \hat{H}}{\partial g_{ab}}|\psi_0}\rangle = \langle{\psi_1|\frac{\partial \hat{H}}{\partial g_{ab}}|\psi_1\rangle}
    \label{eqn: appendix expectations of H derivatives}
\end{equation}
holds at the minimum of the first $L = 0$ excitation mode.
By differentiating Elosing q.~\eqref{eqn: appendix hamiltonian} with respect to $g_{ab}$ and inserting the second quantized representation of the delta interaction we find for an arbitrary state $|\psi_j\rangle$
\begin{equation}
    \begin{split}
        \langle\psi_j|\frac{\partial\hat{H}}{\partial g_{ab}}|\psi_j\rangle &= \langle \psi_j|\sum_{i=1}^{N_b}\delta(\theta_{b,i} - \theta_a )|\psi_j\rangle\\
        &= \sum_{i=1}^{N_b} \sum_{\alpha,\beta,\kappa,\gamma} \Delta^i_{\alpha,\beta,\kappa,\gamma} \langle \psi_j| \hat{b}^\dagger_\alpha\hat{a}^\dagger_\beta\hat{b}_\kappa\hat{a}_\gamma|\psi_i\rangle\\
        &= N_b\int d\theta_a \rho^{(2,j)}_{ba}(\theta_b = \theta_a, \theta_a)\\
    \end{split}
    \label{eqn: appendix H derivative derivation}
\end{equation}
where
\begin{equation}
    \begin{split}
        \Delta^i_{\alpha,\beta,\kappa,\gamma} &= \int d\theta_a \phi_\alpha^*(\theta_{a}) \phi_\beta^*(\theta_a) \phi_\kappa(\theta_{a}) \phi_\gamma(\theta_a).
    \end{split}
    \label{eqn: appendix coefficient of 2nd quantized operator}
\end{equation}
From Eq.~\eqref{eqn: appendix expectations of H derivatives} and~\eqref{eqn: appendix H derivative derivation} and setting the position of the impurity to ${\theta_a = 0}$ it follows that
\begin{equation}
    \rho^{(2,0)}_{ba}(0,0) = \rho^{(2,1)}_{ba}(0,0)
\end{equation}
at the minima of the first ${L = 0}$ excitation mode.

\section{Bogoliubov-de-Gennes Excitations}\label{Appendix: BdG}                                                                                                                          
We calculate the excitation energies for a system of $N_b$ repulsive bosons and a single attractive and infinitely massive impurity ${(M_a\rightarrow \infty)}$ confined to a one-dimensional ring of radius $R$, where the ground state of the system can be described in mean field theory by the Gross–Pitaevskii equation (GPE)
\begin{equation}\label{eqn: BdG Appendix GPE}
    i\hbar \frac{\partial\psi}{\partial t} = \hat{H}_\textrm{MF} \psi.
\end{equation}
Here $\psi$ is the wave function of the bosons and $\hat{H}_\textrm{MF}$ is the MF Hamiltonian defined by             
\begin{equation}\label{eqn: MF Hamiltonian}
    \hat{H}_\textrm{MF}=-\frac{\hbar^2}{2M_bR^2}\frac{\partial ^2}{\partial \theta ^2} +g_{bb}\frac{N_b-1}{N_b R}|\psi|^2+\frac{g_{ab}}{R}\delta(\theta),
\end{equation}
with the same parameters given in Sec.~\ref{Sec: Model} of the main text.
By introducing quasiparticle amplitudes $u_n(\theta)$ and $v_n(\theta)$, we can write the $n^\textrm{th}$ excited state ${\psi}_n(\theta, t)$ as a linear expansion around the ground state $\psi_{0}$ of Eq.~\eqref{eqn: MF Hamiltonian} by using the Bogoliubov ansatz \cite{GAO2020109058}                                   
\begin{equation}\label{eqn: BdG Appendix Bogoliubov ansatz}
    \psi_n = e^{-i\mu_b t}\left[\psi_{0}(\theta)+u_n(\theta)e^{-i\omega_n t}+v_n^*(\theta)e^{i\omega_n^* t}\right].
\end{equation}        
By propagating Eq.~\eqref{eqn: BdG Appendix GPE} in imaginary time, we find the ground state, from which we also calculate the chemical potential $\mu_b$. To arrive 
at the Bogoliubov-de Gennes equations we insert the ansatz, Eq.~\eqref{eqn: BdG Appendix Bogoliubov ansatz}, into the GPE, Eq.~\eqref{eqn: BdG Appendix GPE}, and keep terms linear in the quasiparticle amplitudes. 
This results in the following system of equations for the quasiparticle amplitudes
\begin{equation}
    \mathbf{M}\mathbf{u}_n=\hbar \omega_n \mathbf{u}_n
\end{equation}
with the vector 
\begin{equation}
    \mathbf{u}_n = (u_n,v_n)^T
\end{equation}
and matrix~\cite{GAO2020109058}
\begin{align}                                                                           
    \mathbf{M}=\begin{pmatrix}
        \hat{H}_{MF}-\mu_b + G   & G  \\                           
-G  & -\hat{H}_{MF}+\mu_b - G
\end{pmatrix}                 
\end{align}                          
where $G=g_{bb}(N_b -1)\psi_{0}^2/N_bR$. We diagonalize the matrix $\mathbf{M}$ by standard diagonalization methods to arrive at the BdG excitation energies presented in Figs.~\ref{fig: MF Comparison},~\ref{fig: first excitation for N = 6 g = 0.3 and various M_a },~\ref{fig: BdG modes with increasing N} of the main text.

\section{Reduced Mass Scaling}\label{Appendix: Reduced Mass Scaling}

\begin{figure}[t!]
    \centering
    \includegraphics[width=\columnwidth]{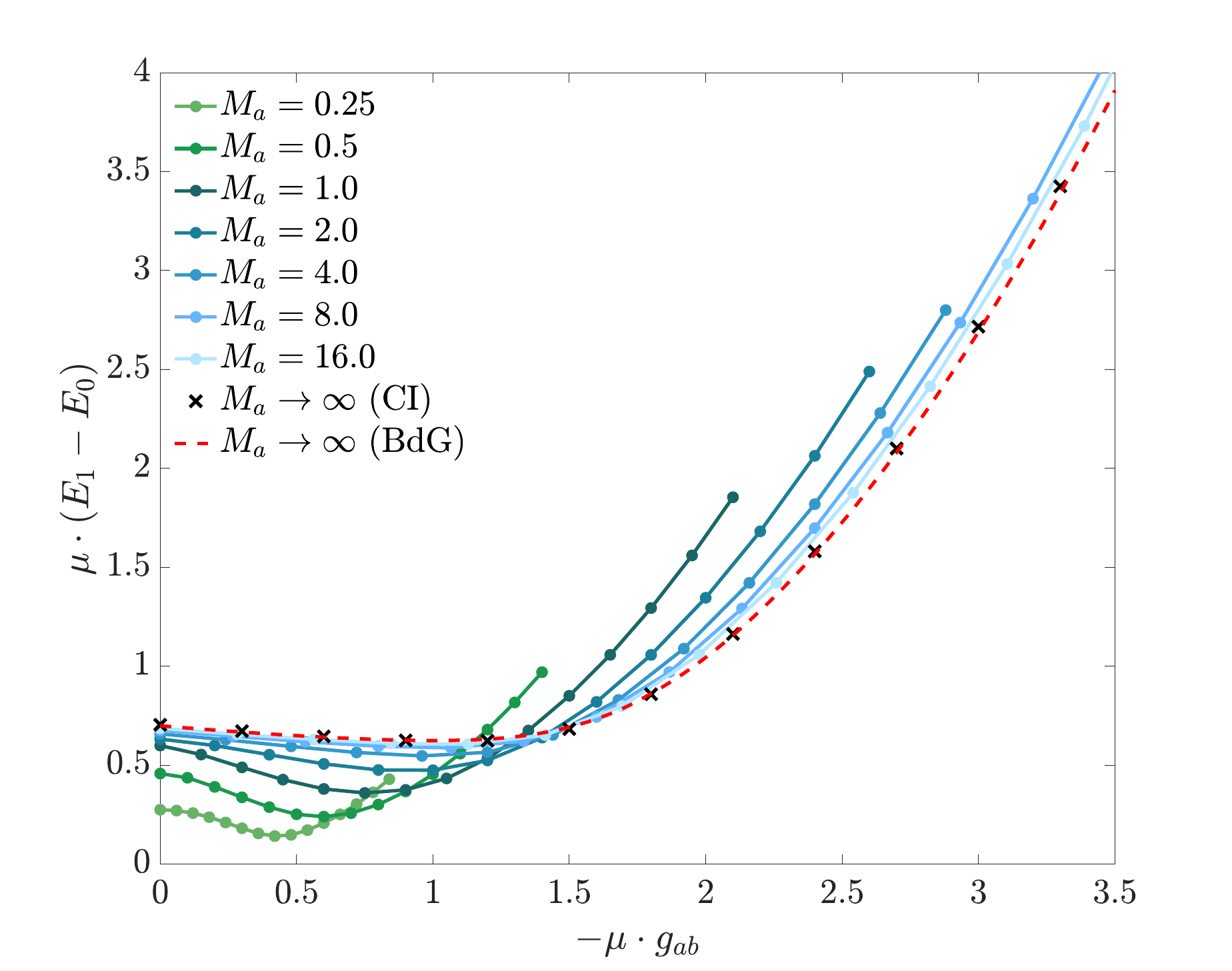}
    \caption{The data presented in Fig.~\ref{fig: first excitation for N = 6 g = 0.3 and various M_a } of the main text, where the energies and interactions have been rescaled with the reduced mass $\mu = M_a/(M_a + 1)$. The first ${L = 0}$ excitation as a function of boson-impurity interaction strength, $\mu g_{ab}$, for a system of ${N_b = 6}$, ${g_{bb} = 0.3}$, and various impurity masses, ${0.25 \leq M_a \leq 16.0}$. The first excitation for the Hamiltonian with an explicitly broken rotational symmetry due to the fixed delta potential $(M_a \rightarrow \infty)$, calculated in the CI framework, is plotted in black crosses. The red dashed line shows the first BdG excitation of the numerical ground state solution to the associated GPE.} 
    \label{fig: first excitation for N = 6 g = 0.3 and various M_a mu scaling}
    \centering
\end{figure}

In Sec.~\ref{Sec: Comparison to Mean-Field} we note that according to mean-field calculations of Ref.~\cite{Brauneis_2022}, the energy of the finite mass impurity system can be related to the energy of the system with a static impurity via the scaling of energies and interaction parameters with the reduced mass, $\mu = M_a/(M_a + 1)$, shown in Eq.~\eqref{eqn: energy scaling with mu}.
Figures~\ref{fig: first excitation for N = 6 g = 0.3 and various M_a mu scaling}~(a) and ~(b) demonstrate a departure from this scaling, indicating the importance of beyond-mean-field effects and the necessity of CI methods.

\begin{figure}[hpt]
    \centering
    \includegraphics[width=\columnwidth]{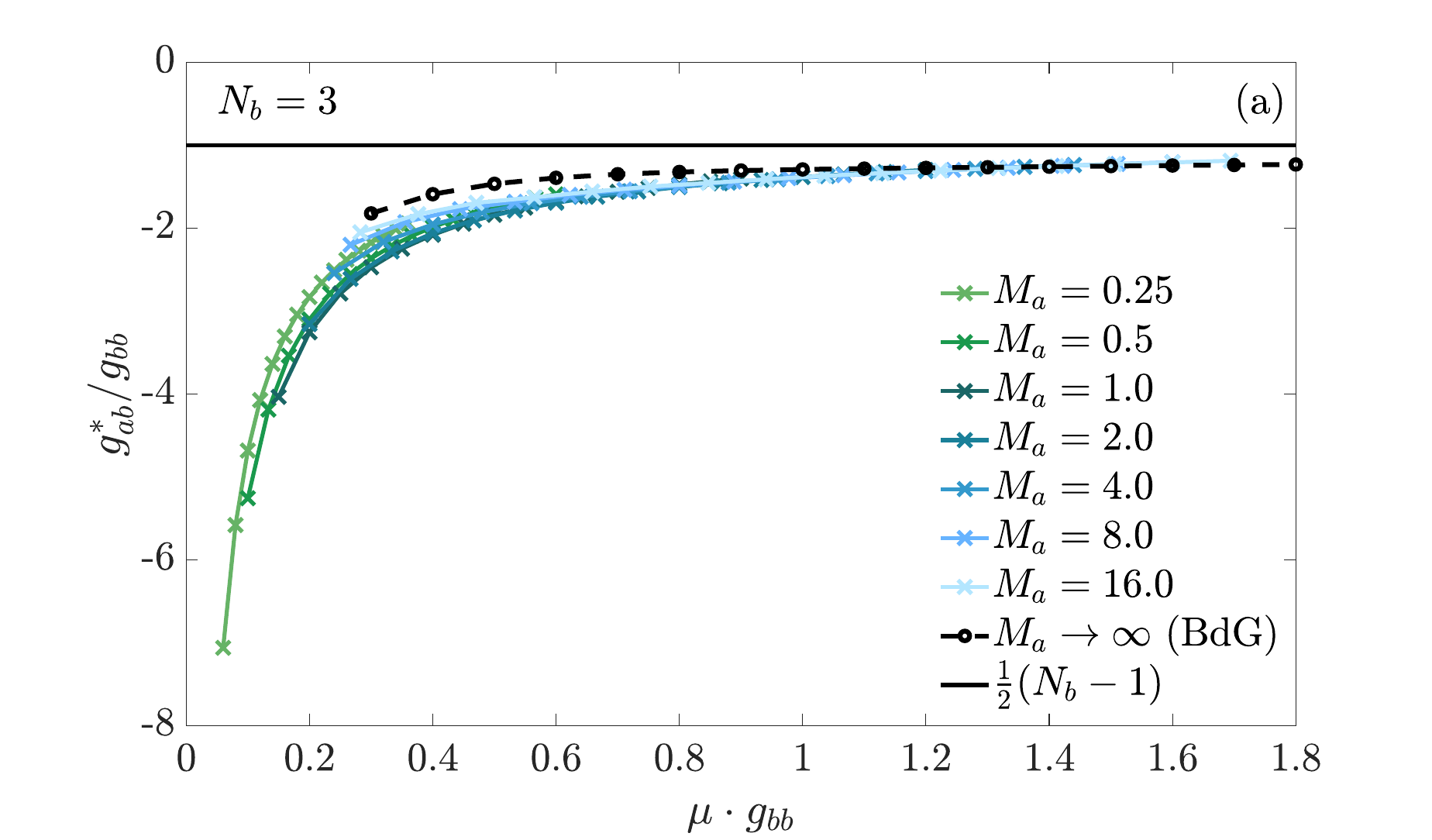}
    \includegraphics[width=\columnwidth]{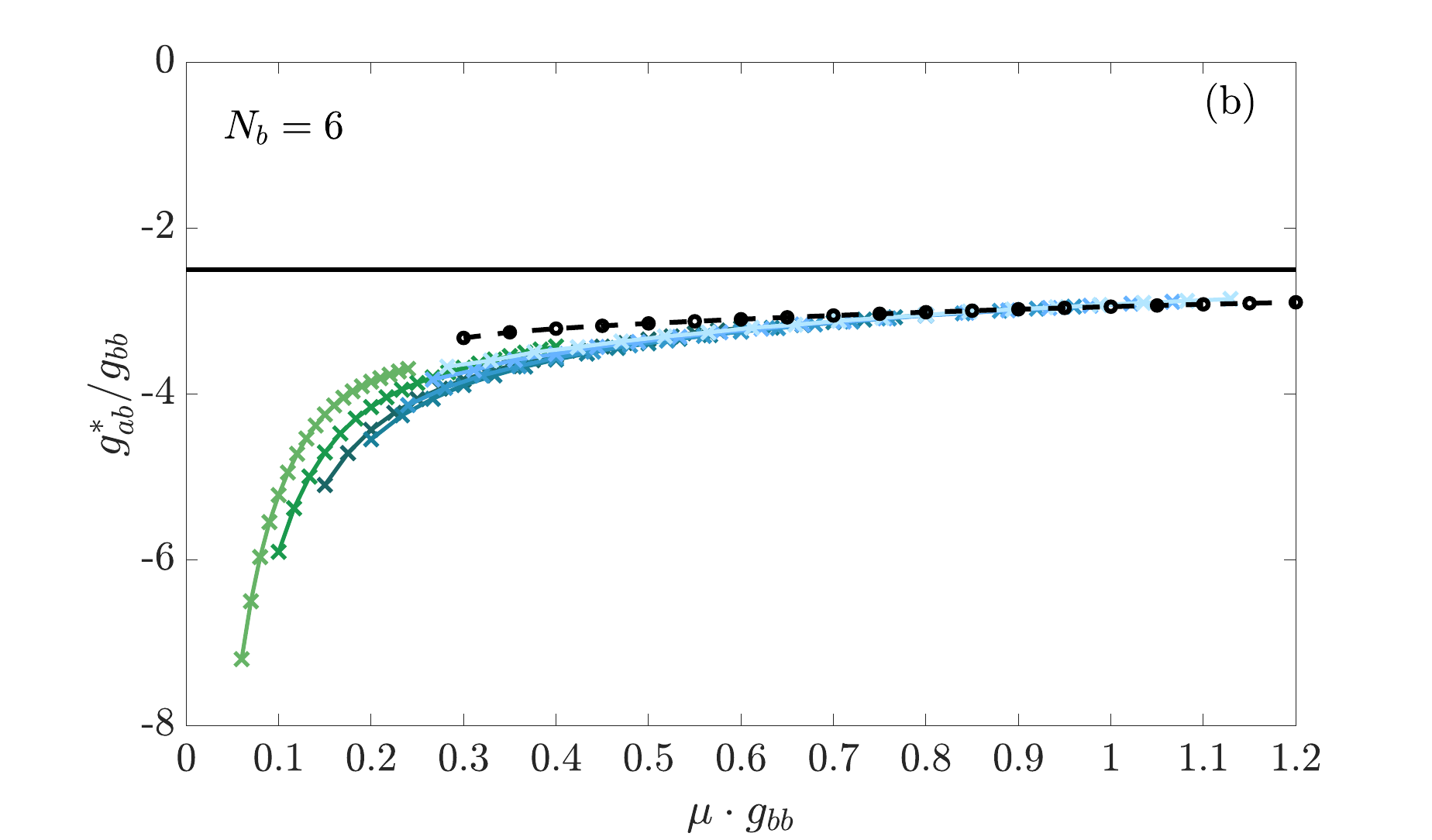}
    \caption{The first $L=0$ excitation mode minima presented in Fig.~\ref{fig: MF Comparison}~(a) and~(b) of the main text, where the interactions have been rescaled with the reduced mass $\mu = M_a/(M_a +1)$.
    The position of the minimum at impurity-boson interaction strength $g_{ab}^*$, relative to the boson-boson repulsion, $g_{bb}$, as a function of $\mu g_{bb}$ for a system of ${N_b =  3}$ (top) and ${N_b = 6}$ (bottom) bosons and various impurity masses, $M_a$. The solid black line indicates ${\frac{1}{2}(N_b - 1)}$, the MF prediction of the position of the phase transition from Eq.~\eqref{eqn: MF Ncrit condition} of the main text~\cite{Brauneis_2022}. The dot-dashed line indicates the position of the HA minima found from the BdG excitation to the numerical ground state solution of the corresponding GPE for a system of $N_b$ bosons with an infinitely massive (fixed delta potential) impurity. } 
    \label{fig: MF Comparison mu scaling}
    \centering
\end{figure}

\clearpage
%\bibliography{Bib} 

%\bibliographystyle{apsrev4-1}


\end{document}